\documentclass[aps,amsmath,preprint,epsf,superscriptaddress,nofootinbib]{revtex4-1}
\usepackage{graphicx}
\usepackage{bm}
\usepackage{color}
\usepackage{amsmath,amssymb}	
\usepackage{hyperref}
\usepackage{setspace}
\usepackage{longtable}
\usepackage{booktabs}
\usepackage{comment}
\usepackage{array}
\usepackage{cancel}
\usepackage{enumitem}

\def\slashchar#1{\setbox0=\hbox{$#1$} 
\dimen0=\wd0 
\setbox1=\hbox{/} \dimen1=\wd1 
\ifdim\dimen0>\dimen1 
\rlap{\hbox to \dimen0{\hfil/\hfil}} 
#1 
\else 
\rlap{\hbox to \dimen1{\hfil$#1$\hfil}} 
/ 
\fi}

\def\beq{\begin{eqnarray}}
\def\eeq{\end{eqnarray}}

\begin{document}
\newcolumntype{Y}{>{\centering\arraybackslash}p{23pt}} 


\preprint{IPMU19-0044}
\preprint{TU-1076}

\title{\large \bf  Muon $g-2$ in Split-Family SUSY in light of LHC Run II}

\author{Masahiro Ibe}
\email[e-mail: ]{ibe@icrr.u-tokyo.ac.jp}
\affiliation{ICRR, The University of Tokyo, Kashiwa, Chiba 277-8582, Japan}
\affiliation{Kavli IPMU (WPI), UTIAS, The University of Tokyo, Kashiwa, Chiba 277-8583, Japan}
\author{Motoo Suzuki}
\email[e-mail: ]{m0t@icrr.u-tokyo.ac.jp}
\affiliation{ICRR, The University of Tokyo, Kashiwa, Chiba 277-8582, Japan}
\affiliation{Kavli IPMU (WPI), UTIAS, The University of Tokyo, Kashiwa, Chiba 277-8583, Japan}
\author{Tsutomu T. Yanagida}
\email[e-mail: ]{tsutomu.tyanagida@ipmu.jp}
\affiliation{T.D.Lee Institute and School of Physics and Astronomy, Shanghai Jiao Tong University, Shanghai 200240, China}
\affiliation{Kavli IPMU (WPI), UTIAS, The University of Tokyo, Kashiwa, Chiba 277-8583, Japan}
\author{Norimi Yokozaki}
\email[e-mail: ]{yokozaki@truth.phys.tohoku.ac.jp}
\affiliation{Department of Physics, Tohoku University, Sendai, Miyagi 980-8578, Japan}

\date{\today}
\begin{abstract}
The Split-Family supersymmetry is a model in which the sfermion masses of the first two generations are in $\mathcal{O}(100\text{--}1000)$\,GeV while that of the third one is in $\mathcal{O}(10)$\,TeV. With such a hierarchical spectrum, the deviation of the muon $g-2$  and the observed Higgs boson mass are explained simultaneously.  In this paper, we revisit the Split-Family SUSY model in light of the updated LHC constraints. We also study the  flavor changing neutral current problems in the model.  As we will show, the problems do not lead to stringent constraints  when the Cabibbo-Kobayashi-Maskawa matrix is the only source of the flavor mixing. We also study how large flavor mixing in the supersymmetry breaking parameters is allowed.

\end{abstract}

\maketitle

\section{Introduction}
\label{sec:introduction}
The standard model (SM) of particle physics is now complete by the discovery of the Higgs boson with a mass around $125$\,GeV~\cite{Aad:2012tfa,Chatrchyan:2012xdj, Aad:2015zhl,Khachatryan:2016vau,ATLAS:2018doi}. In the minimal  supersymmetric SM (MSSM)~(see \cite{Nilles:1983ge} and references therein), the measured Higgs boson mass
 can be explained when the masses of the superpartners of the top quark (the stop) are in $\mathcal{O}(10\text{--}100)$\,TeV~\cite{Okada:1990vk,Okada:1990gg,Ellis:1990nz,Haber:1990aw,Ellis:1991zd}. No evidence of the super particles at the LHC experiments also suggests that their masses are in  a multi-TeV range.

On the other hand, the measured value of the muon anomalous magnetic moment  $a_\mu^{\rm exp}$~\cite{Hagiwara:2011af,Keshavarzi:2018mgv} deviates from the SM prediction $a_\mu^{\rm SM}$~\cite{Bennett:2006fi,Roberts:2010cj} at about $3.7 \sigma$,%
\footnote{Davier $et.al.$ have reported a deviation of $3.6~\sigma$ level~\cite{Davier:2010nc}. 
The deviation has increased~\cite{Jegerlehner:2018zrj}  by the recent more precise measurement of the fine-structure constant from atomic interferometry with Cesium133~\cite{Parker:2018}. See~\cite{Lusiani:2018tcd} for the future prospect of the muon $g-2$ measurement.}
\beq
{\mit  \Delta}  a_\mu \equiv a_\mu^{\rm exp}-a_\mu^{\rm SM}=(27.06\pm 7.26)\times 10^{-10}.
\eeq
This discrepancy can be explained in the MSSM when such as the smuons and the electroweakinos are in $\mathcal{O}(100)$\,GeV~\cite{Lopez:1993vi,Chattopadhyay:1995ae,Moroi:1995yh}.  

In~\cite{Ibe:2013oha}, the Split-Family supersymmetry (SUSY) model has been proposed to explain the observed Higgs boson and the muon $g-2$ deviation simultaneously. There, the sfermions of the first two generations are in $\mathcal{O}(100\text{--}1000)$\,GeV while that of the third one is in $\mathcal{O}(10)$\,TeV.%
\footnote{For other simple possibilities and models in the MSSM, see Ref.~$e.g.$ ~\cite{Evans:2011bea,Evans:2012hg,Ibe:2012qu,Sato:2012bf,Akula:2013ioa,Bhattacharyya:2013xba,Bhattacharyya:2013xma,Chakrabortty:2013voa,Evans:2013uza,Mohanty:2013soa,Babu:2014lwa,Gogoladze:2014cha,Iwamoto:2014ywa,Ajaib:2015ika,Chowdhury:2015rja,Gogoladze:2015jua,Harigaya:2015kfa,Harigaya:2015jba,Gogoladze:2016jvm,Yin:2016shg,Yanagida:2017dao,Wang:2018vrr}.} 
Such a hierarchical SUSY spectrum is motivated by the Yukawa hierarchy.

In this paper, we revisit the Split-Family SUSY model in light of the updated LHC constraints.%
\footnote{For model independent study, See~\cite{Ajaib:2015yma,Tran:2018kxv}.} As we will see, almost the entire region which explains the muon $g-2$ within $2\sigma$ is excluded for the universal gaugino mass. We also show that the collider constraints can be evaded for the non-universal gauino masses while explaining the muon $g-2$.

 We also study the FCNC problems in the  Split-Family SUSY model. As we will explain, the precise construction of the model requires careful treatment of the family basis, which generically leads to sizable  SUSY contributions to the flavor changing neutral currents (FCNC). To see such effects, we first discuss the case where the Cabibbo-Kobayashi-Maskawa (CKM) matrix is the only source of the flavor mixing. As we will see, the FCNC constraints are not so stringent in that case.  We also discuss how large flavor mixing in the supersymmetry breaking parameters are allowed. 
 
The organization of this paper is as follows. In Sec.~\ref{sec:model}, we review the Split-Family SUSY model. In Sec.~\ref{sec:pheno}, we update the constraints from the collider experiments. In Sec.~\ref{sec:fcnc}, we discuss the FCNC constraints on the model. In Sec.~\ref{sec:btauuni}, we discuss the successful bottom-tau unification as a bonus feature of the model. The final section is devoted to our conclusion. 

\section{Split-Family SUSY Model}
\label{sec:model}
The basic idea of the Split-Family model is to assume the hierarchical soft masses,
\beq
\label{eq:mass}
m_{\rm soft}^2=
\left(
\begin{array}{ccc}
m_0^2 & 0 & 0 \\
0 & m_0^2 & 0 \\
0 & 0 & m_3^2 
\end{array}
\right),
\eeq
where the masses of the first two generation sfermions, $m_0^2$, are in $m_0=\mathcal{O}(100-1000)$\,GeV and that of the third generation sfermions, $m_3^2$, is in $m_3=\mathcal{O}(10)$\,TeV. This structure may be related to the hierarchy of the Yukawa couplings. For example, an extra-dimensional setup can lead to the aligned hierarchy between the soft masses and the Yukawa couplings, where only the third generation resides on the brane close to those of the Higgs brane and the SUSY breaking brane (See $e.g.$~\cite{Gabella:2007cp}). It is also possible to achieve the aligned hierarchy if the first two generations are pseudo-Nambu-Goldstone multiplets of some broken global symmetry  (See $e.g.$~\cite{Kugo:1983ai,Yanagida:1985jc,Evans:2013uza}).

The precise construction of the Split-Family model requires careful treatment of the family basis. If the soft masses are universal, for example, the general Yukawa couplings in the superpotential do not lead to the SUSY FCNC contributions, since the soft breaking parameters are proportional to the unit matrix in any family basis. On the other hand, for the non-universal soft masses, the general Yukawa couplings result in the non-zero SUSY contributions to the flavor mixing. Since we have assumed the rather light sfermions, those mixing leads to unacceptably large FCNC processes. 

In this paper, we put a phenomenological requirement that the soft masses in Eq.\,(\ref{eq:mass}) are realized in the family basis where the Yukawa couplings in the superpotential are almost diagonal.%
\footnote{We also assume that the hierarchy of the soft masses and the Yukawa couplings are aligned.} These assumptions are implicitly made, for example, in the models~\cite{Gabella:2007cp,Kugo:1983ai,Yanagida:1985jc,Ibe:2013oha} to avoid too large FCNC. 
It should be noted, however, that there are unavoidable SUSY FCNC contributions even under these assumptions due to the effects of the CKM matrix.
To demonstrate those effects, we study the following two scenarios.
\begin{itemize}
\item The CKM matrix is the only source for the flavor mixing (the minimal mixing scenario)
\item Small flavor mixing comes from the supersymmetry breaking parameters (the small mixing scenario)
\end{itemize}
In the following, we call the first scenario the minimal mixing scenario and the second one the small mixing scenario. The minimal mixing scenario gives us a demonstration of how large SUSY FCNC contributions are expected by the effect of the non-universality of the soft masses in Eq.\,(\ref{eq:mass}).

\subsubsection*{Minimal Mixing Scenario}
Let us start from the minimal mixing scenario. The superpotential with general Yukawa matrices is given by,
\beq
W_{\rm LEPTON}&=&f_E^{ij}L_i\bar E_jH_d,\\
W_{\rm QUARK}&=&f_U^{ij}Q_i\bar U_j H_u+f_D^{ij}Q_i\bar D_j H_d.
\eeq
Here, $L,~\bar E,~Q,~\bar U,~\bar D$ are chiral superfields of the left-handed leptons, the left-handed anti- electrons, the left-handed quarks, the left-handed up-type anti-quarks, and the left-handed down type anti-quarks, respectively. The coefficients $f_E,~f_U,~f_D$ are the $3\times 3$ complex matrices. The subscripts $i,~j$ run from one to three and each one is contracted. 

As we assume that the soft masses in Eq.\,(\ref{eq:mass}) are defined in the family basis where the Yukawa couplings are diagonal, the lepton Yukawa matrix is given by the diagonal form,
\beq
\label{eq:slepton}
W_{\rm LEPTON}=\hat{f}_E^{ii}L^{[e]}_i\bar E_i^{[e]}H_d,
\eeq
where the subscript $[e]$ shows the chiral superfields in the diagonalized basis, $\hat{f}_E$ is a diagonal and real-positive matrix.%
\footnote{Even if we include the PMNS effect, the lepton flavor mixing is not so large for the model with the degenerate right-handed masses $M_R\lesssim 10^{10}$\,GeV (See the appendix~\ref{sec:lfvmns} for more details). Thus, we ignore the finite neutrino masses in the superpotential.} Note that the subscript $i$ in Eq.\,(\ref{eq:slepton}) runs from one to three, which corresponds to the charged lepton generation. 

 On the other hand, the up- and down-type quark Yukawa matrices are not diagonalized at the same time due to the CKM matrix. To demonstrate the minimal mixing scenario, we take a simple family basis, for example,
\beq
\label{eq:SCKM}
W=\hat{f}^{ii}_U Q^{[d]}_i {\bar U}^{[u]}_i H_u+(V_{\rm CKM}^*\hat{f}_D)^{ij} Q_i^{[d]} {\bar D}^{[d]}_j H_d. 
\eeq
Here, $\hat{f}_U$ and $\hat{f}_D$ are diagonal and real-positive matrices, $V_{CKM}$ is the CKM matrix, the superscript $*$ is a complex conjugate of a matrix, and the superscript $T$ denotes the transpose of a matrix. The subscripts $i,~j$ in Eq.\,(\ref{eq:SCKM}), run from one to three, which corresponds to the quark generation. In this basis, the down-type Yukawa couplings are not diagonal due to the CKM matrix. We may consider another simple family basis where the CKM matrix appears in the up-type Yukawa couplings%
\footnote{That is, we may take the basis, 
\beq
\label{eq:SCKMu}
W=(V_{\rm CKM}^T\hat{f}_U)^{ij} Q_i {\bar U}_j H_u+(\hat{f}_D)^{ii} Q_i{\bar D}_i H_d,
\eeq 
while the soft masses are given by Eq.\,(\ref{eq:mass}).
} (See Sec.~\ref{sec:fcnc} for more details).

\subsubsection*{Small Mixing Scenario}
The small mixing scenario is defined in the following way.
We first take the family basis specified in Eq.\,(\ref{eq:slepton}) and Eq.\,(\ref{eq:SCKM}). Then, we introduce the small flavor mixing matrix $V_{\rm mix}$ so that the mass matrices are given by 
\beq
\label{eq:massmix}
m_{\rm soft}^2=V^{\dagger}_{\rm mix}
\left(
\begin{array}{ccc}
m_0^2 & 0 & 0 \\
0 & m_0^2 & 0 \\
0 & 0 & m_3^2 
\end{array}
\right)V_{\rm mix}.
\eeq
In the small mixing scenario, we assume that $V_{\rm mix}$ is close to the unit matrix with the mixing angles of the order of the CKM angles.
See Sec.~\ref{sec:fcnc} for more details.

\section{Phenomenology of Split-Family SUSY model}
\label{sec:pheno}
In this section, we update the favored region for the muon $g-2$ and the Higgs boson mass in Ref.~\cite{Ibe:2013oha} in light of the current LHC data. The SUSY contributions to the FCNC are discussed in the next section.

\subsection{Parameter Choice At Input Scale}
 The free parameters in our analysis are,
\beq
m_0,~m_3,~m_{H_u}^2,~m_{H_d}^2,~{\rm tan}\beta,~M_1,~M_2,~M_3.
\eeq
Here, $m^2_{H_u}~(m_{H_d}^2)$ is the squared mass of the up-type (down-type) Higgs doublet, ${\rm tan}\beta$ is defined by the ratio of the vacuum expectation value of $H_u$ to that of $H_d$, and $M_1,~M_2$, and $M_3$ are the gaugino mass parameters of the bino, the wino, and the gluino, respectively. 
We assume the same $m_0^2$ and $m_3^2$ for the squarks and the sleptons for simplicity.%
\footnote{These assumptions are motivated by the $SU(5)$ GUT. }

We take the above free parameters at the GUT scale $M_{\rm GUT}=10^{16}\,{\rm GeV}$. By solving the renormalization group equations, we obtain the physical parameters. Notice that the Higgsino mass parameter $\mu$ is chosen to achieve electroweak symmetry breaking consistently (See Eq.\,(\ref{eq:mu})). It should be also noted that the following arguments are not sensitive to the choice of $m_{H_u}^2$ and $m_{H_d}^2$ as long as the electroweak symmetry breaking is successful.%
\footnote{When the required value of $\mu$ is as small as $m_0$, the Higgsino also contributes to the muon $g-2$~\cite{Ibe:2013oha}.} For simplicity, we take
\beq
\label{eq:hmassrelation}
m_{H_u}^2=m_{H_d}^2=m_0^2,
\eeq
in the following analysis. 

For $m^2_{H_u}=m_{H_d}^2=\mathcal{O}(m_0^2)$, radiative corrections from the stop mass squared leads to a relatively large $|m^2_{H_u}|$ at the weak scale and it requires a large $|\mu|$. That is, the value of $|\mu|$ is determined by the minimization condition of Higgs potential,
\beq
\label{eq:mu}
|\mu|^2&=&-m^2_{H_u}+\frac{1}{{\rm tan}\beta^2}(m^2_{H_d}-m^2_{H_u})-\frac{1}{2}m_Z^2+\mathcal{O}\left(\frac{1}{{\rm tan}\beta^4}\right)\simeq-m_{H_u}^2,
\eeq
for ${\rm tan}\beta \gg 1$. Here, $m_Z$ is the mass of the $Z$ boson, $m_Z\simeq 91$\,GeV. From the numerical analysis, $|m_{H_u}^2|$ at the electroweak scale is $\mathcal{O}(10)^2\,{\rm TeV}^2$ in the parameter space of our interest, and thus $|\mu|=\mathcal{O}(10)\,$TeV. 

To explain the muon $g-2$, we require that the masses of the electroweakinos are also in $\mathcal{O}(100-1000)$\,{\rm GeV}. We consider two cases, the universal gaugino mass and
the non-universal gaugino masses. In the case of the universal gaugino mass, we take
\beq
M_{1/2}=M_1=M_2=M_3=\mathcal{O}(100-1000)\,{\rm GeV}.
\eeq
For the non-universal gaugino masses, we take
\beq
M_1\simeq M_2=\mathcal{O}(100-1000)\,{\rm GeV},~M_3\gtrsim -\mathcal{O}(1)\,{\rm TeV}.
\eeq
The reason of the choice of the sign of $M_3$ will be explained in Sec.~\ref{sec:btauuni}.

The SUSY contributions to the muon $g-2$ are proportional to ${\rm tan}\beta$ and we focus on
\beq
{\rm tan}\beta\gtrsim 40.
\eeq
The large ${\rm tan}\beta$ is also advantageous to explain the observed Higgs boson mass with $m_3=\mathcal{O}(10)$\,TeV. As we will also discuss, the large ${\rm tan}\beta$ is important to achieve the bottom-tau unification (See Sec.~\ref{sec:btauuni}).

Throughout this paper, we assume that the SUSY breaking parameters do not have CP violating phases and the only source of the CP violation in the following analysis comes from the CP phase in the CKM matrix. Under these assumptions, we take the convention of  $\mu>0$ without loss of generality (See the following section for more details).

 \subsection{SUSY Contributions To Muon $g-2$}
 Here, we show the parameter space to explain the muon $g-2$. In the MSSM, the relevant one loop contributions to the muon $g-2$ come from the diagrams with the smuon/neutralino or the muon-type sneutrino/chargino loops~\cite{Cho:2011rk}. In our scenario, the one-loop bino-smuon diagram dominates the SUSY contributions to the muon $g-2$, the SUSY contribution is proportional to the $\mu$ parameter.
The Higgsino contributions are, on the other hand, suppressed due to their heaviness.

In our analysis, we use the package {\tt SPheno-4.0.3}%
\footnote{We slightly modify {\tt SPheno-4.0.3}  to calculate the SUSY spectrum on the basis in Eq.\,(\ref{eq:SCKM}).} to calculate the low energy spectrum from the input parameter at the GUT scale. In the code, the renormalization group equations are solved including the flavor mixing parameters, which are relevant for analyses in the next section. We also  use the package {\tt FeynHiggs2.14.3}~\cite{Frank:2006yh,Heinemeyer:1998np,Heinemeyer:1998yj,Hahn:2009zz,Degrassi:2002fi} to compute the muon $g-2$ and the Higgs mass from the low energy spectrum obtained by {\tt SPheno}.%
\footnote{We added the two loop corrections for the large ${\rm tan}\beta$ given in~\cite{Marchetti:2008hw} for the SUSY contributions to the muon $g-2$. Although we use {\tt FeynHiggs} to calculate the muon $g-2$, the following results are not changed even if we use {\tt SPheno}.}

\begin{figure}[tbp]
\begin{center}
  \begin{minipage}{.4\linewidth}
  \includegraphics[width=\linewidth]{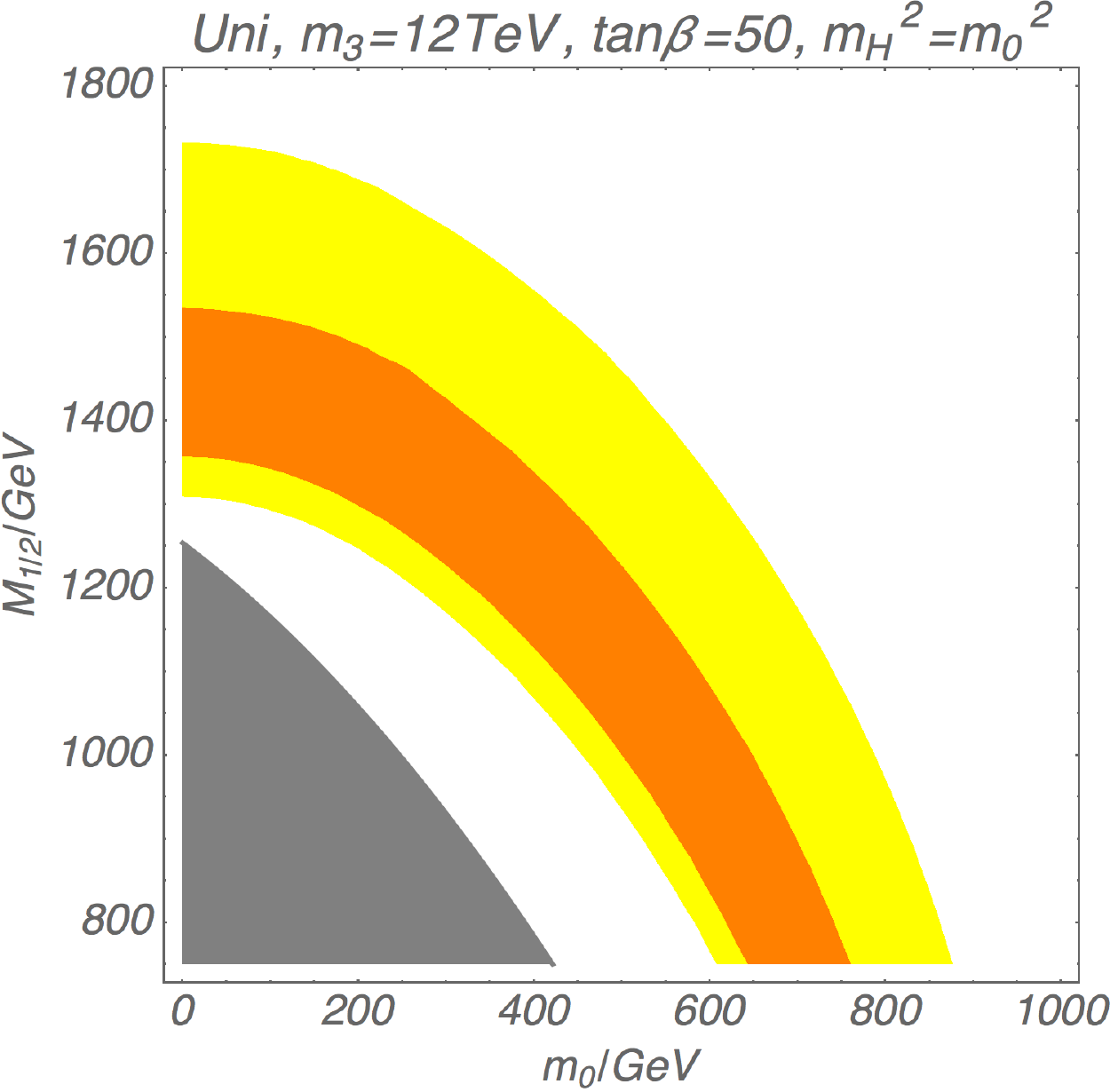}
 \end{minipage}
  \hspace{1cm}  
  \begin{minipage}{.4\linewidth}
  \includegraphics[width=\linewidth]{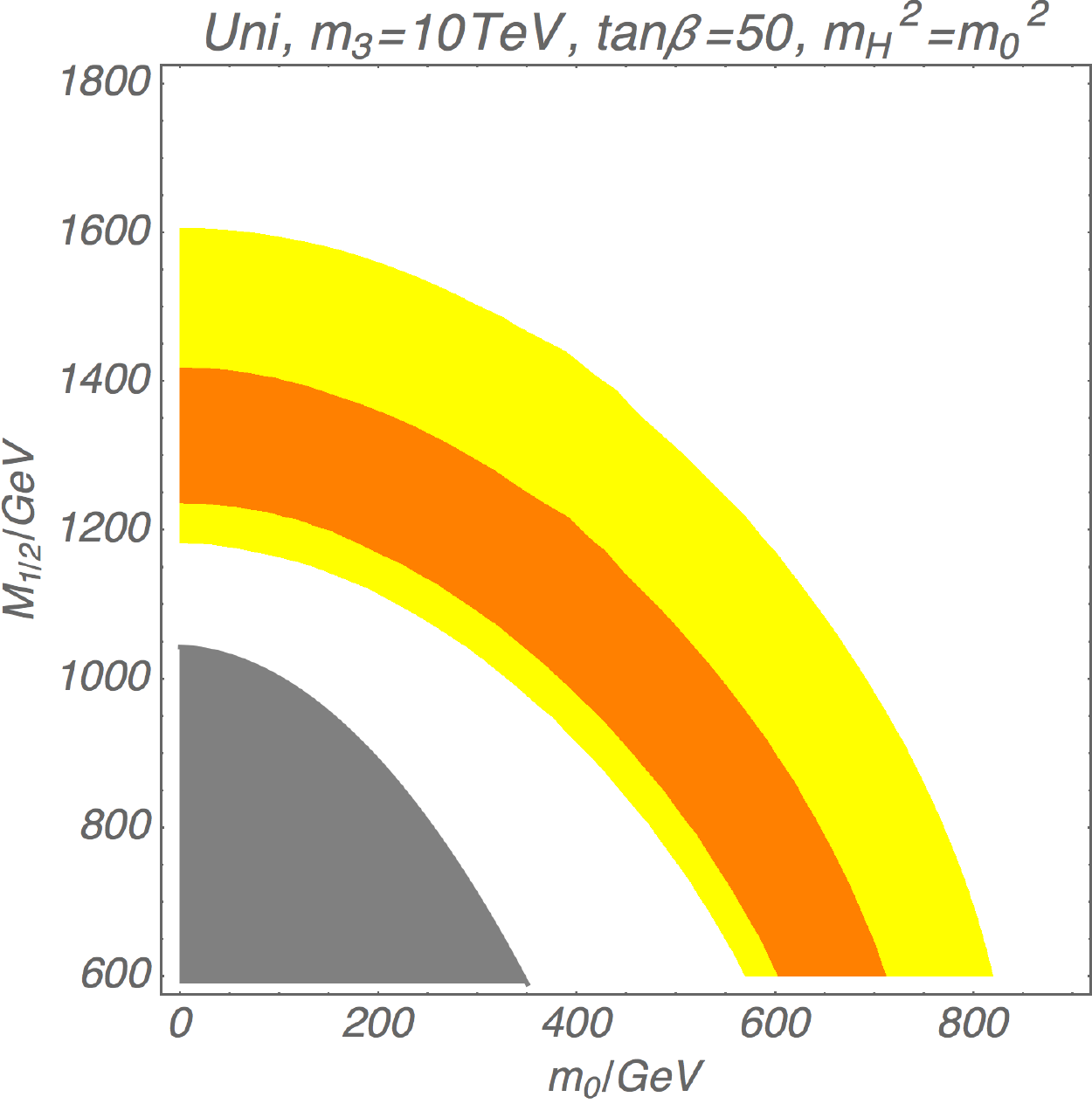}
 \end{minipage}
 \end{center}
  
\caption{\sl \small
The plots for the muon $g-2$  on the $(m_0\text{--}M_{1/2})$ plane for given $m_3$, ${\rm tan}\beta$, and $m_H^2=m^2_{H_{u,d}}$. We take ${\rm tan}\beta=50$ and $m_3=12\,{\rm TeV}~\text{or}~10\,{\rm TeV}$.  In the orange (yellow) region, ${\mit  \Delta}  a_{\mu}$ is explained within $1\sigma~(2\sigma)$. The gray shaded region is excluded by the negative slepton masses.  The Higgs boson mass is consistent with the observed Higgs boson mass in the favored regions for ${\mit  \Delta}  a_{\mu}$.
}
\label{fig:uniorg}
\end{figure}

\subsubsection{Universal Gaugino Mass at the GUT Scale}
 \label{sec:uni}
\begin{figure}[htbp]
\begin{center}
  \begin{minipage}{.4\linewidth}
  \includegraphics[width=\linewidth]{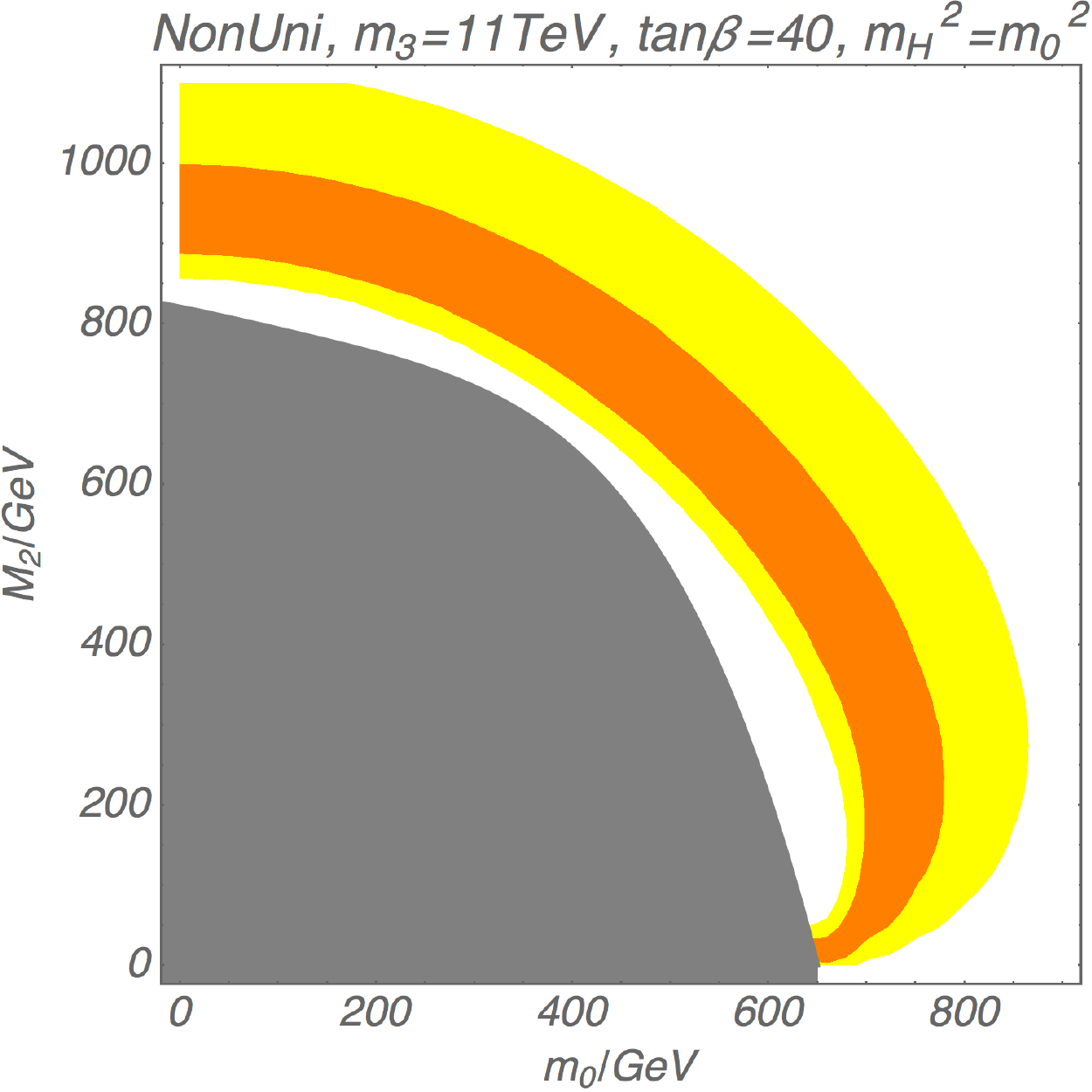}
 \end{minipage}
   \hspace{1cm} 
  \begin{minipage}{.4\linewidth}
  \includegraphics[width=\linewidth]{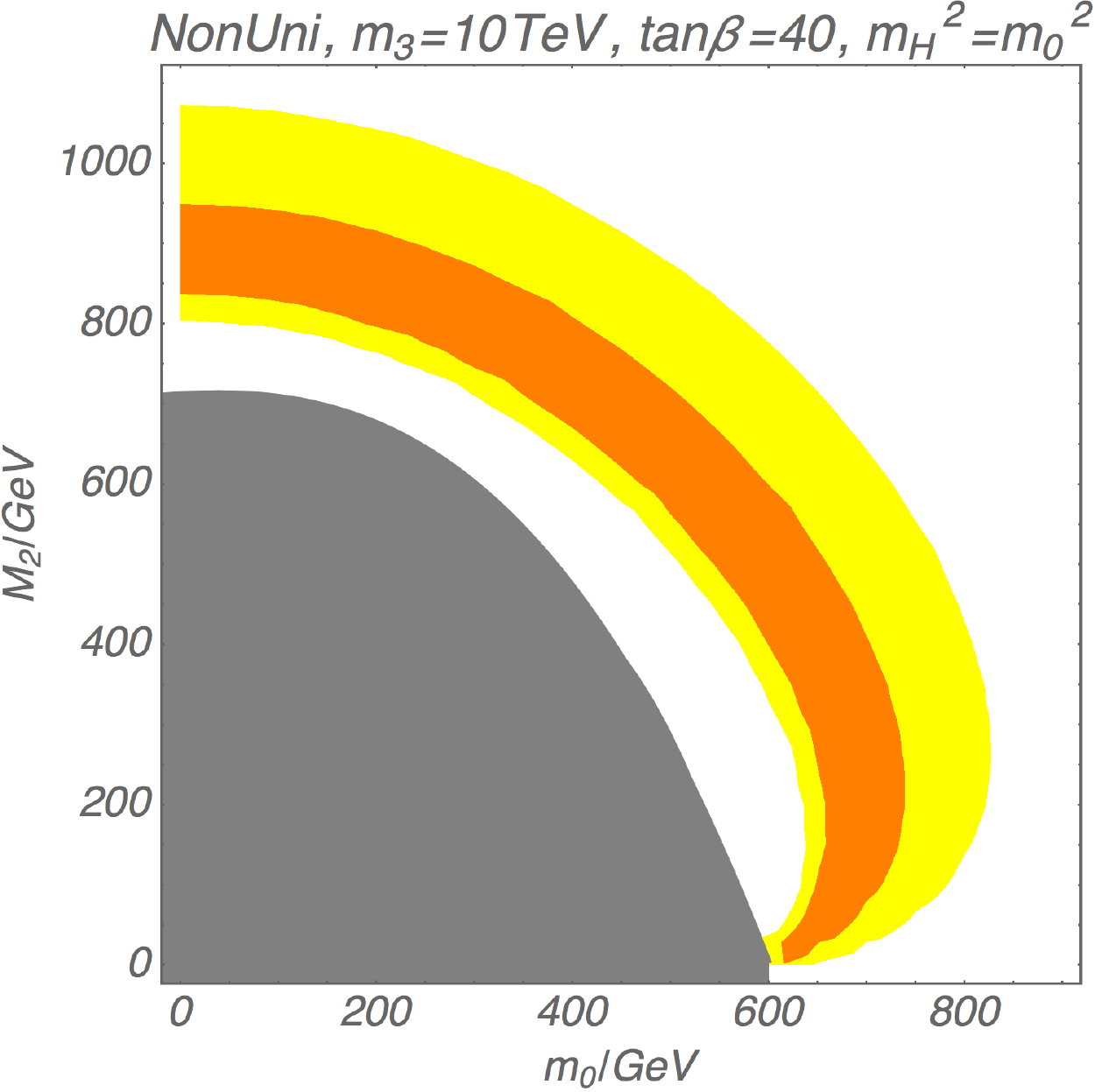}
 \end{minipage}
 \end{center}
  
\caption{\sl \small
The plots for the muon $g-2$ on the $(m_0-M_{2})$ plane with the non-universal gaugino masses $M_1\simeq 1.7M_2$ at the GUT scale.  We take $M_1=1.725\times M_2~(M_1=1.74\times M_2)$ and the gluino mass parameter $M_3=-2.6\,{\rm TeV}~(-2.5\,{\rm TeV})$ at the GUT scale for $m_3=11\,{\rm TeV}~(10\,{\rm TeV})$. The color codes are the same as Fig.~\ref{fig:uniorg}. The Higgs boson mass is consistent with the observed one in the favored regions. }
\label{fig:nonuniSorg}
\end{figure}

For the universal gaugino mass, we search for the region to explain the observed muon $g-2$ on the $m_0\text{--}M_{1/2}$ plane for $m_3=12\,{\rm TeV}$ or $10\,{\rm TeV}$. In Fig.~\ref{fig:uniorg}, the observed muon $g-2$ is explained in the orange (yellow) regions within $1\sigma$~($2\sigma$). The gray shaded regions are excluded by the tachyonic masses of the sleptons in the first two generations due to the large two-loop renormalization group effects from the third generation masses. In most of the favored parameter space, the SM Higgs mass is in $124{\rm -}126\,{\rm GeV}$ for the central value of the measured top quark mass $m_{\rm top}=173.1\pm 0.9$\,GeV~\cite{Tanabashi:2018oca}. That is consistent with the observed Higgs boson mass $m_H=125.18\pm0.16$\,GeV~\cite{Tanabashi:2018oca} within the theoretical uncertainty.%
\footnote{The theoretical uncertainty of the Higgs mass is about $2-3$\,GeV~\cite{Athron:2016fuq}. We have also checked that the Higgs boson mass is consistent with the observed value within the theoretical error in Fig.~\ref{fig:nonuniSorg},~\ref{fig:uni},~\ref{fig:nonuniS}, and Fig.~\ref{fig:nonunibt}.} In the left figure, the measured muon $g-2$ is explained for slightly larger $M_{1/2}$ and $m_0$ compared with the right one. This is because the $\mu$ parameter is larger for the larger $m_3$, with which the SUSY contribution to the muon $g-2$ is slightly enhanced.

\subsubsection{Non-Universal Gaugino Masses at the GUT Scale}
For the non-universal gaugino masses, we also study the parameter space explaining the muon $g-2$ on the $m_0\text{--}M_2$ plane. In Fig.~\ref{fig:nonuniSorg}, we plot the orange (yellow) shaded regions predicting the observed muon $g-2$ within $1\sigma~(2\sigma)$. The gray shaded regions are excluded by the tachyonic masses of the sleptons. Here, the ratio of the gaugino masses are fixed to be $M_1=1.725\times M_2$ ($M_1=1.74\times M_2$) and $M_3=-2.6\,{\rm TeV}$ ($-2.5\,{\rm TeV}$) for $m_3=11$\,TeV ($10$\,TeV) at the GUT scale, respectively. The motivations of these choices will be explained in Sec.~\ref{sec:nonuniS}. Again, the SM Higgs boson mass is consistent with the observed Higgs boson mass in the favored parameter space. As in the case of the universal gaugino mass, the muon $g-2$ is explained by slightly larger $m_0$ and $M_2$ for the larger $m_3$ when we compare both the figures.

\subsection{Collider Constraints}
Now, let us discuss the collider constraints. We will discuss cosmology in the next subsection.

\begin{figure}[htbp]
\begin{center}
  \begin{minipage}{.4\linewidth}
  \includegraphics[width=\linewidth]{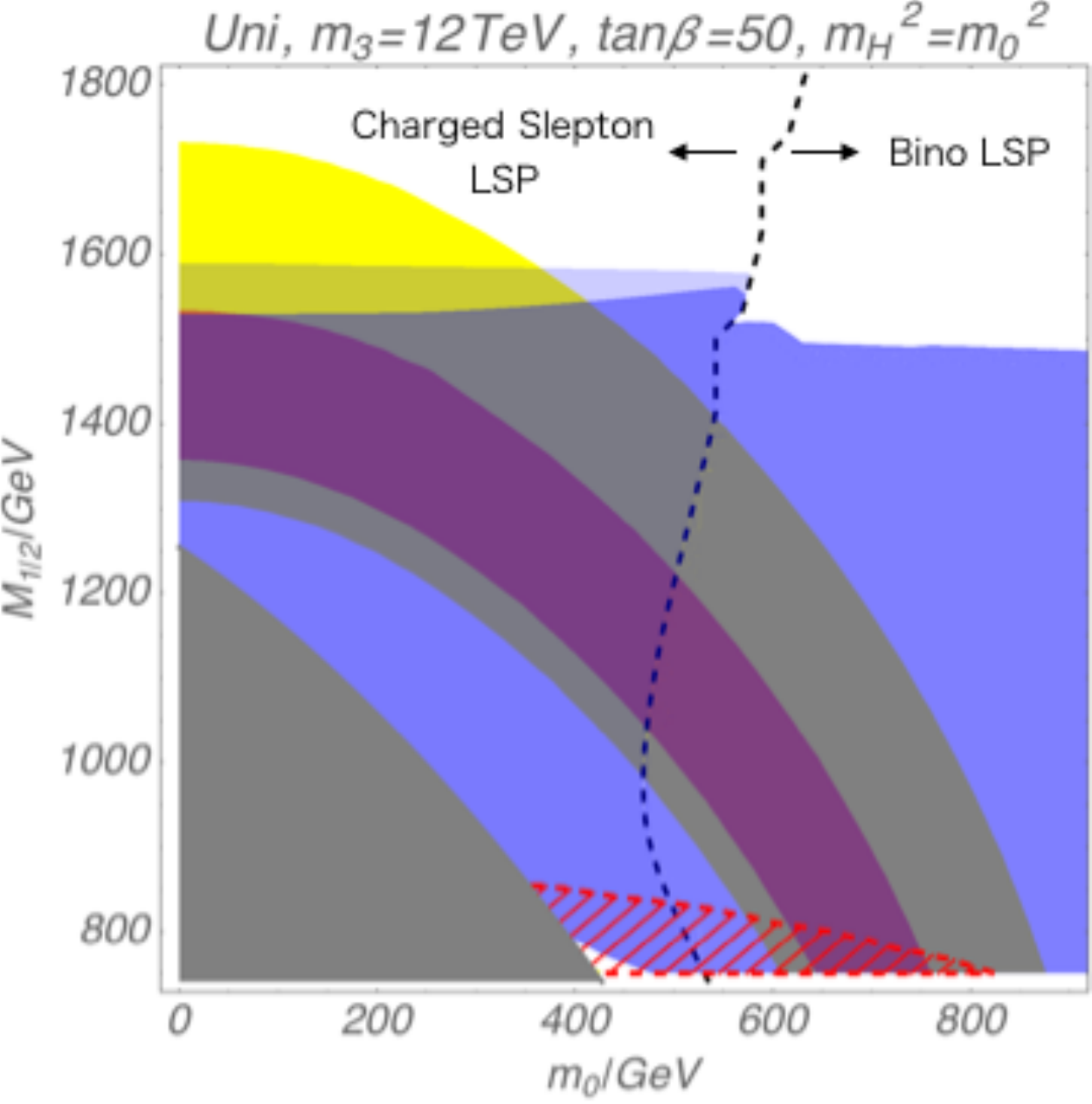}
 \end{minipage}
  \hspace{1cm}  
  \begin{minipage}{.4\linewidth}
  \includegraphics[width=\linewidth]{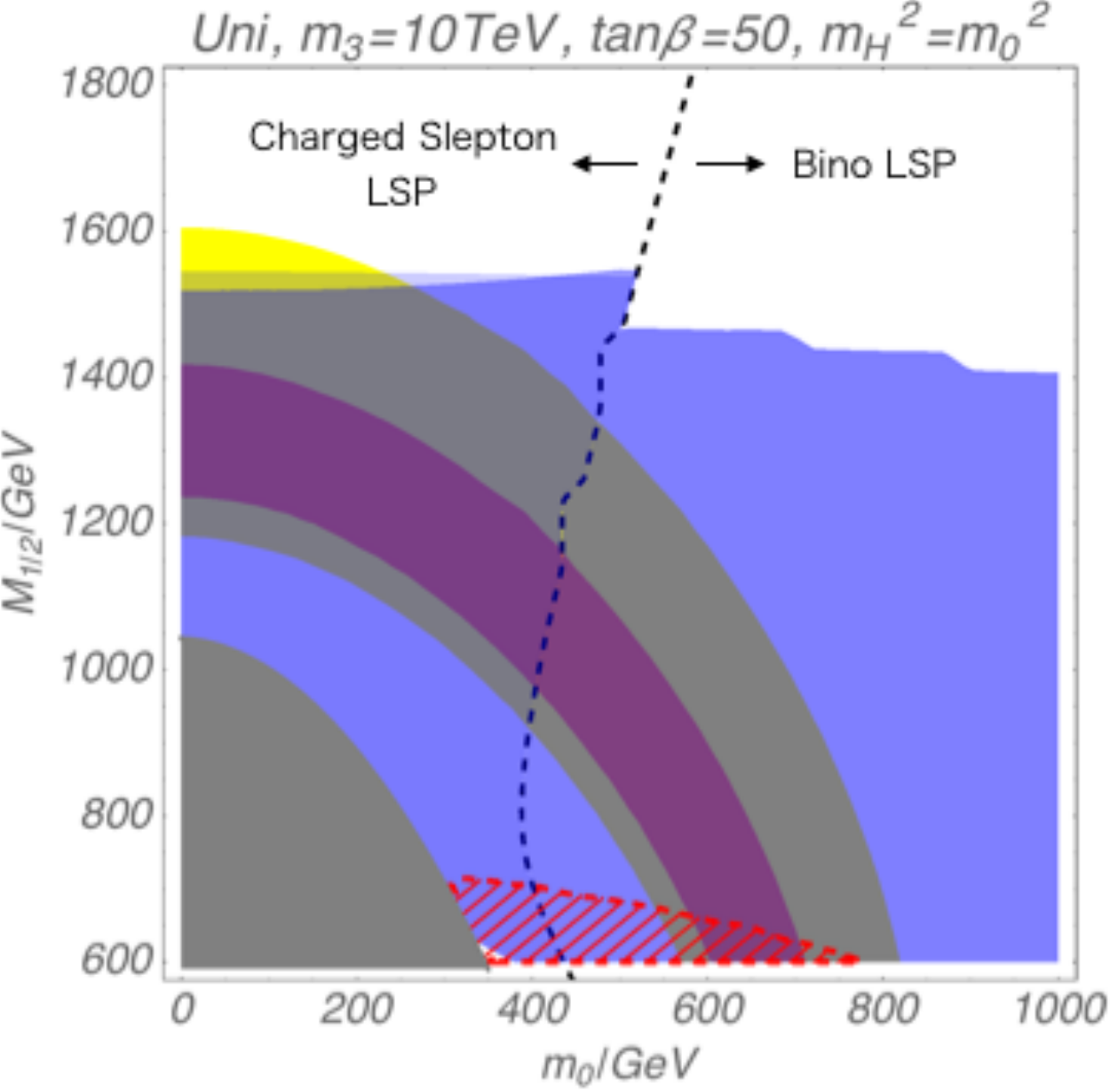}
 \end{minipage}
 \end{center}
  
\caption{\sl \small
Summary plots for the collider constraints. The same with Fig.~\ref{fig:uniorg} for the parameter space and the color codes. On the right (left) side of the black dashed line, the bino (the charged slepton) is the LSP. The blue shaded regions correspond to the $95\%$\,CL exclusion limit in~\cite{Aaboud:2017vwy} (See the text for the difference between the blue and lighter blue regions). The red hatched regions are excluded by $\epsilon_K$ ($90$\% CL) (See Sec.~\ref{sec:fcnc}).
}
\label{fig:uni}
\end{figure}

\begin{table}[htbp]
\caption{\sl 
A sample point in the universal gaugino mass case. $m_{\rm  gluino}$, $m_{\tilde{Q}}$, $m_{\tilde{e}_L}~(m_{\tilde{\mu}_L})$, $m_{\tilde{e}_R}~(m_{\tilde{\mu}_R})$, $m_{\tilde{\chi}^1_0}$, $m_{\tilde{\chi}^2_0}$, $m_{\tilde{\chi}_1^\pm}$ denote the masses of the gluino, the lightest squark, the lightest almost left handed selectron (smuon), the lightest almost right-handed selectron (smuon), the lightest neutralino, the next to the lightest neutralino, the lightest chargino, respectively. It should be noted that $R$-parity violation is required for the decay of the LSP before the Big-Bang Nucleosynthesis (BBN). Then, we also need to introduce a dark matter candidate in the model. (See the next subsection.)
}
\small{
\begin{center}
\begin{tabular}{c|c}
$m_0,~m_3$ & $0\,{\rm GeV},~12\,{\rm TeV}$ \\ 
$M_{1/2}$ & $1650\,{\rm GeV}$ \\ 
${\rm tan}\beta$ & $50$ \\ \hline\hline
$m_{\rm Higgs}$ & $124.6\,{\rm GeV}$\\
$a_{\mu}$ & $1.48\times 10^{-9}$\\ 
LSP & \text{charged slepton}\\
${\mit  \Delta} M_K$ & $-3.0\times 10^{-21}\,{\rm GeV}$ \\
$\epsilon_K$ & $-2.6\times 10^{-7}$ \\
$ {\mit  \Delta} M_D$ & $1.5\times 10^{-17}\,{\rm GeV}$ \\ 
\hline
$\mu$ & $9751~{\rm GeV}$ \\
$m_{\rm gluino}$  & $3619\,{\rm GeV}$ \\
$m_{\tilde{Q}}$    &  $2730 \,{\rm GeV}$ \\
$m_{\tilde{e}_L}(m_{\tilde{\mu}_L})$ & $486\,{\rm GeV}$ \\
$m_{\tilde{e}_R}(m_{\tilde{\mu}_R})$ & $942\,{\rm GeV}$\\
$m_{\chi_0^1}$ & $766\,{\rm GeV}$\\
$m_{\chi_0^2}$ & $1360\,{\rm GeV}$\\
$m_{\chi_1^\pm}$ & $1360\,{\rm GeV}$\\
\end{tabular}
\end{center}}
\label{tab:uni}
\end{table}%

\subsubsection{Universal Gaugino Mass at the GUT Scale}
\label{sec:collideruni}
The bino is the lightest SUSY particle  (LSP) on the right sides of the black dashed lines in Fig.~\ref{fig:uni}. In this case, the parameter space with the gluino and the squarks lighter than about $2.6$\,TeV is excluded by the search for multi-jets plus missing transverse momentum at ATLAS $13$\,TeV using $36$\,fb$^{-1}$~\cite{Aaboud:2017vwy}. We show the excluded regions as the blue shaded ones in Fig.~\ref{fig:uni}, which correspond to the $95\%$\,CL limits. 

A charged slepton is the LSP on the left side of the black dashed line in Fig.~\ref{fig:uni}. 
Here, we assume that the LSP has a short lifetime by $R$-parity violation so that the scenario is consistent with cosmology. However, the size of the $R$-parity violation is limited from above not to wash out the baryon asymmetry made by baryogenesis (such as thermal leptogenesis). As a result, the charged slepton LSP is expected to be stable inside the detectors (See $e.g.$~\cite{Endo:2009cv}). 
 The heavy stable charged particles searches in~\cite{CMS:2016ybj} put upper limits on the production cross section of the SUSY particles, which is converted to the constraints on the mass parameters by using the cross-section given in~\cite{Borschensky:2014cia}.

  In Fig.~\ref{fig:uni}, the blue shaded regions on the left of the black dashed lines are excluded by the constraints on the heavy stable charged particle ($95\%$\,CL).%
\footnote{There are two kinds of blue regions (the blue and the lighter blue regions). The blue plus light blue shaded regions are excluded here. We will explain this difference soon.}
In the figure, almost entire region favored by the muon $g-2$ is excluded except for a tiny region near $(m_0,\,M_{1/2})=(0\,{\rm GeV},\,1.7\,{\rm TeV})$ for $m_3=12\,{\rm TeV}$ in the case of the universal gaugino mass.%
\footnote{The muon $g-2$ can be explained for a larger $M_{1/2}$ for a larger ${\rm tan}\beta$. In this case, however,  the CP-odd neutral Higgs scalar becomes tachyonic. 
}  In Tab.~\ref{tab:uni}, we show a sample spectrum which evades the constraints while explains the muon $g-2$ within $2\sigma$ in the charged slepton LSP scenario.

In the analysis for the charged slepton LSP, the electroweakino productions are the dominant SUSY production modes, where we use the production cross section given in~\cite{Fuks:2012qx,Fuks:2013vua}. For comparison, we also show the constraints assuming the SUSY production cross section for the degenerate squarks and gluino (the lighter blue shaded regions). In the actual spectrum, the gluino is heavier than the lightest squark in most of the favored parameter space, and hence, the production cross section of the colored SUSY particles~\cite{Borschensky:2014cia} is smaller than the degenerated case.
 As the figure shows, the constraints via the colored SUSY particle production are at most comparable or weaker than those from the electroweakino productions for the heavy stable charged particle searches.%
 \footnote{For the searches of multi-jets plus missing transverse energy which is relevant for the bino search, the colored SUSY production plays the dominant roles.}

\begin{figure}[tbp]
\begin{center}
  \begin{minipage}{.4\linewidth}
  \includegraphics[width=\linewidth]{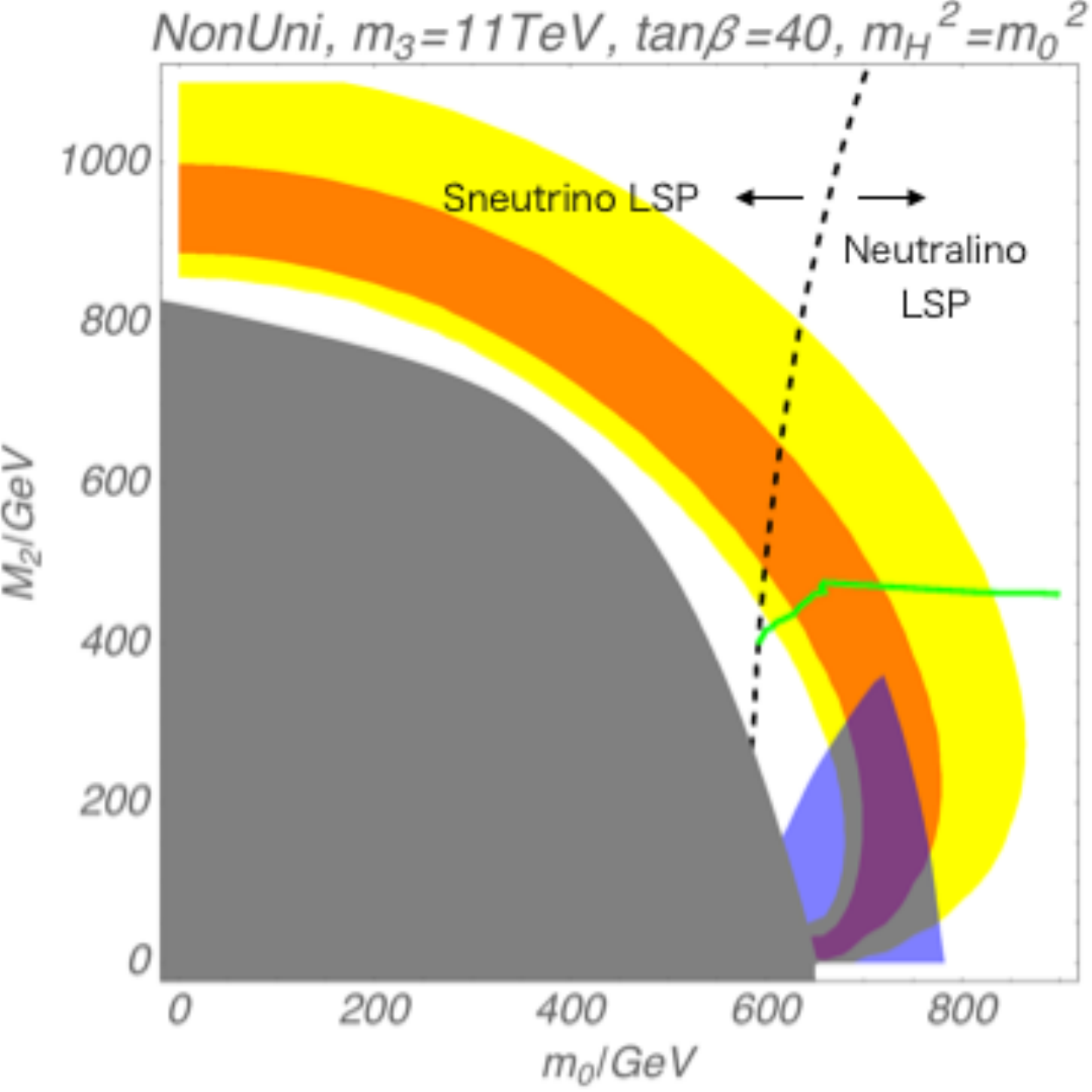}
 \end{minipage}
   \hspace{1cm} 
  \begin{minipage}{.4\linewidth}
  \includegraphics[width=\linewidth]{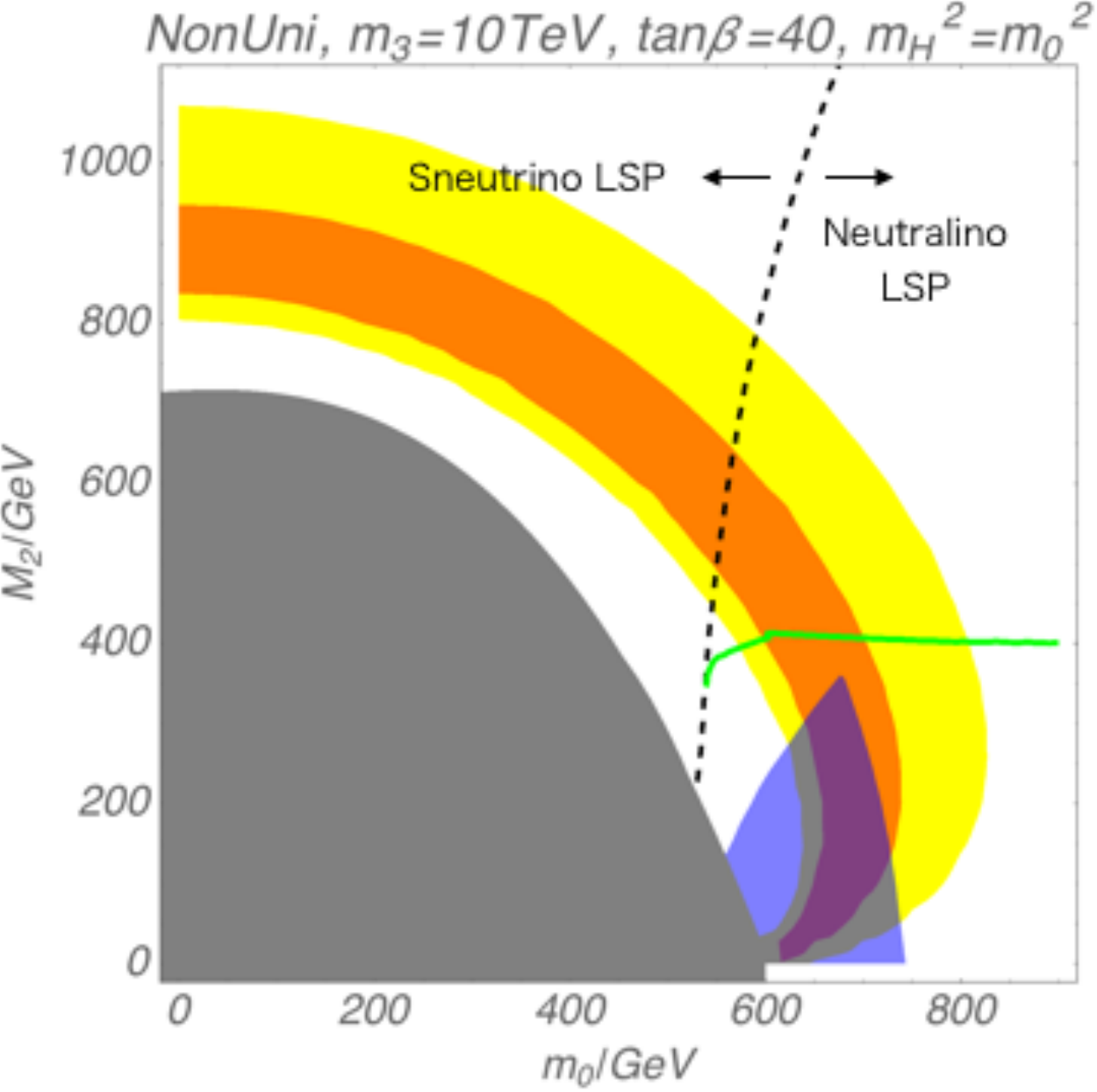}
 \end{minipage}
 \end{center}
  
\caption{\sl \small
The summary plots for the collider constraints and the dark matter candidate. The same with Fig.~\ref{fig:nonuniSorg} for the parameter space and the color codes. On the right (left) side of the black dashed line, the neutralino (the sneutrino) is the LSP. The blue shaded regions are excluded from the collider searches at $95\%$\,CL. On the green line, the observed current dark matter abundance $\Omega h^2\simeq0.12$ is achieved due to the bino-wino coannihilation. 
}
\label{fig:nonuniS}
\end{figure}

\begin{table}[ht]
\caption{\sl 
A sample parameter point in the case of the non-universal gaugino masses. The neutralino is the LSP. $\Omega h^2$ and $\sigma^{\rm SI}$ denote the current thermal relic abundance of the bino and the spin-independent bino-nucleon cross-section.} 
\small{
\begin{center}
\begin{tabular}{c|c}
$m_0,~m_3$ & $650\,{\rm GeV},~11\,{\rm TeV}$ \\ 
$M_{1},~M_{2},~M_3$ &~$1.725\times 480\,{\rm GeV},~ 480\,{\rm GeV},~-2.6\,{\rm TeV}$ \\ 
${\rm tan}\beta$ & $40$ \\ \hline\hline
$m_{\rm Higgs}$ & $124.6\,{\rm GeV}$\\
$a_{\mu}$ & $2.69\times 10^{-9}$\\ 
LSP & bino \\
$\Omega h^2$ &  $0.119$\\
$\sigma^{\rm SI}$ & $2.4\times 10^{-14}\,{\rm pb}$ \\
${\mit  \Delta}  M_K$ & $-2.0\times 10^{-21}\,{\rm GeV}$\\
$\epsilon_K$ & $1.2\times 10^{-7}$ \\
${\mit  \Delta}  M_D$ & $1.5\times 10^{-17}\,{\rm GeV}$ \\ \hline
$\mu$ & $9108\,{\rm GeV}$\\ 
$m_{\rm gluino}$ & $5529\,{\rm GeV}$\\
$m_{\tilde{Q}}$ & $4442\,{\rm GeV}$\\
$m_{\tilde{e}_L}(m_{\tilde{\mu}_L})$ & $481\,{\rm GeV}$\\
$m_{\tilde{e}_R}(m_{\tilde{\mu}_R})$ & $635\,{\rm GeV}$\\
$m_{\chi_0^1}$ & $404\,{\rm GeV}$\\
$m_{\chi_0^2}$ & $478\,{\rm GeV}$\\
$m_{\chi_1^\pm}$ & $478\,{\rm GeV}$\\
\end{tabular}
\end{center}}
\label{tab:nonuni1}
\end{table}%

 \subsubsection{Non-Universal Gaugino Masses at the GUT Scale}
 \label{sec:nonuniS}
In the case of the non-universal gaugino masses, the constraints from the collider searches can be weakened by several reasons. For the bino LSP cases, for example, the production cross section of the colored SUSY particles which are relevant for multi-jets plus missing transverse energy search is reduced if the gluino and squarks are heavy. The constraints from the heavy stable charged particle searches can be also evaded since the sneutrino can be lighter than the charged sleptons. In the following, we again take $M_3=-2.6\,{\rm TeV}$ or $-2.5\,{\rm TeV}$ at the GUT scale, which suppresses the colored SUSY particle production cross section. The choice of the sign of $M_3$ will be relevant in the discussion in Sec.~\ref{sec:btauuni}. We also take $M_1=1.725\times M_2$ or $M_1=1.74\times M_2$ with which the sneurinos are lighter than the charged sleptons in the favored parameter space.

On the right side of the black dashed line in Fig.~\ref{fig:nonuniS}, the neutralino is the LSP. As we have mentioned, there is no stringent collider constraint because the squarks and the gluinos become heavy by a rather large $|M_3|$. It should be also noted that the constraints on the missing transverse momentum from the electroweakino production are far less relevant due to the rather degenerate electroweakino spectrum, $(m_{\chi_1^\pm}-m_{\chi_0^1})/m_{\chi_1^\pm}\lesssim 30\% $~\cite{Aaboud:2018sua,Sirunyan:2017lae}. 

Eventually, the searches for the missing transverse energy with the charged leptons put the most stringent constraints on the neutralino LSP region. In the figure, the blue shaded regions in Fig.~\ref{fig:nonuniS} are excluded by the constraint in \cite{Aaboud:2018jiw} ($95\%$\,CL).%
\footnote{The constraint from the latest result in~\cite{ATLAS:2019cfv} is not stringent, where it is assumed that the slepton decays into a lepton and a neutralino with a $100$\% branching ratio. In our scenario, a slepton also decays into a chargino and a neutrino. Furthermore, a charged lepton from a chargino decay becomes soft. Thus, the constraint is weakened.}
 In Tab.~\ref{tab:nonuni1}, we show a sample spectrum which evades the constraints while explains the muon $g-2$ within $1\sigma$ in the neutralino LSP scenario.

On the left side of the black dashed line in the favored parameter space in Fig.~\ref{fig:nonuniS}, the sneturino is the LSP  due to a relatively small $M_2$ compared with $M_1$. Because of the heavy colored SUSY particles and the degenerate electroweakinos, this region is less constrained by multi-jet plus missing energy searches~\cite{Aaboud:2017vwy,Aaboud:2018sua,Sirunyan:2017lae}. The search for the missing transverse energy with the charged leptons are also less sensitive to this region. In the sneutrino LSP region, however, the constraints from the direct detection of dark matter are stringent if the sneutrino LSP is stable (See the following section for more detail discussion).

\subsection{Cosmology}
As we have seen above, most parameter region favored by the muon $g-2$ has been excluded for the universal gaugino mass by the LHC searches. In this subsection, we discuss cosmology focusing on the non-universal gaugino masses.

In most of the neutralino LSP region, the bino is the dominant component of the lightest neutralino.
In general, the bino LSP is disfavored from a too large thermal relic abundance due to its small annihilation cross section.%
\footnote{In some of the neutralino LSP region, the wino is the LSP. There, we have confirmed that the constraints from the direct detection are negligible due to the small thermal relic abundance of the wino.}
This problem can be evaded for our particular choice of the gaugino mass parameters,
\beq
\label{eq:degeneratem12}
M_1(M_{\rm GUT})\simeq 1.7\times M_2(M_{\rm GUT}),
\eeq
which leads to the rather degenerate physical bino and wino masses. In this case, the co-annihilation between the bino and the wino is efficient~\cite{BirkedalHansen:2002am,Baer:2005jq,ArkaniHamed:2006mb,Harigaya:2014dwa}%
\footnote{See the other neutralino coannihilation with other light sparticles $e.g$ a stop~\cite{Boehm:1999bj}, stau~\cite{Ellis:1998kh}, gluino~\cite{Profumo:2004wk}.}, which makes the bino abundance consistent with the observed dark matter density.

In Fig.~\ref{fig:nonuniS}, we draw the green line on which the thermal relic abundance of the bino corresponds to the observed dark matter density. To calculate the thermal relic abundance of dark matter (and its cross-section with nucleons), we use the package {\tt MicrOMEGAs\_5.0.4}~\cite{Belanger:2018mqt}. For the choice of the gaugino mass relation in Eq.\,(\ref{eq:degeneratem12}), we find that the bino mass around $400$\,GeV can explain the observed dark matter density. By tuning the ratio between $M_2$ and $M_1$ further, the dark matter density can be explained for the bino mass up to around $700$\,GeV within the parameter region favored by the muon $g-2$. 

As the bino is the LSP, the direct detection cross is highly suppressed. In fact, we find,
\beq
\sigma_{\rm SI} \lesssim 10^{-10}\left(\frac{m_{\rm DM}}{100{\rm\,GeV}} \right){\rm pb} 
\eeq
where $m_{\rm DM}$ denotes the dark matter mass. This cross section is much smaller than the current constraint~\cite{Akerib:2016vxi,Cui:2017nnn,Aprile:2018dbl}. 

For the sneutrino LSP, on the other hand, the relic abundance is much smaller than the observed dark matter density ($\Omega h^2 =10^{-2}$) due to its large annihilation cross section in the parameter region favored by the muon $g-2$. Even with such a small relic abundance, however, the sneutrino LSP contribution to dark matter has been excluded by the direct detection experiments since it has a large scattering cross section with the nucleons, $\sigma_{\rm SI} \simeq \mathcal{O}(10^{-5})\,{\rm pb}$.

As a result, we find that tiny $R$-parity violation is required in the case of the sneutrino LSP as in the case of the charged slepton LSP (See the previous discussion in Sec.~\ref{sec:collideruni}). In those cases, we need dark matter candidate other than the LSP.

\newpage
\section{FCNC constraints in Split-Family SUSY model}
\label{sec:fcnc}
.

In the Split-Family model, there is a non-trivial enhancement of the FCNC by the non-universality of the sfermion masses. In this section, we investigate the FCNC constraints on the model for the minimal and small mixing scenarios defined in Sec.~\ref{sec:model}. In the minimal mixing scenario, the CKM matrix is the only source of the flavor mixing. We show that the FCNC constraints on the minimal scenario are not stringent. For the small mixing scenario, we demonstrate how large flavor mixing in the supersymmetry breaking parameters is allowed.

\subsection{Experimental FCNC limits}
Let us first summarize the FCNC constraints relevant to the mixing parameters.

\subsubsection*{Meson mixing}
The CP-violating parameter in the neutral kaon system $\epsilon_K$ gives the stringent constraint on the squark flavor mixing. The measured value of $\epsilon_K$~\cite{Tanabashi:2018oca} is
\beq
|\epsilon_K|^{\rm ex}=2.228(11)\times 10^{-3}  (90\%\,CL).
\eeq
For the theoretical prediction of SM, we adopt~\cite{Jang:2017ieg},
\beq
|\epsilon_K|^{\rm SM,in}=2.05(18)\times 10^{-3},
\eeq
which is derived by using the QCD sum-rule.%
\footnote{We do not use the theoretical prediction on $\epsilon_K$ from the lattice QCD~\cite{Jang:2017ieg}.} 

The mass difference between the long-lived Kaon and the short-lived Kaon, ${\mit  \Delta}  M_K=m_{K^0_L}-m_{K^0_S}$, also puts the constraints on the mixing parameters. The measured ${\mit  \Delta}  M_K$~\cite{Tanabashi:2018oca} is given by
\beq
{\mit  \Delta}  M_K^{\rm exp}=3.483(6)\times 10^{-15}\,{\rm GeV}.
\eeq
The SM prediction of ${\mit  \Delta}  M_K$, however, has a large uncertainty due to the unknown long-distance effects (see $e.g.$ Ref~\cite{Grinstein:2017pvg}). Thus, in this paper, we will just compare ${\mit  \Delta}  M_K^{\rm exp}$ with the order of the magnitude of  SUSY contributions. The neutral $D$-meson oscillation may also put the constraints. The observed mass difference  \cite{Tanabashi:2018oca} is
\beq
|{\mit  \Delta}  M_D^{\rm ex}|=0.63^{+0.27}_{-0.29}\times 10^{-14}\,{\rm GeV}.
\eeq
The theoretical calculation of ${\mit  \Delta}  M_D$ has large uncertainties as in the case of ${\mit  \Delta}  M_K$. Again, we only evaluate the order of the magnitude of the SUSY contribution with the $|{\mit  \Delta}  M_D^{\rm ex}|$.

 \subsubsection*{Lepton Flavor Violation}
For the slepton flavor mixings, the stringent constraint comes from the search for the decay $\mu^+ \rightarrow e^+ \gamma$ at the MEG experiment~\cite{Mori:2016vwi}. The upper limit on the branching ratio of this process is
\beq
\mathcal{B}(\mu^+\rightarrow e^+\gamma) < 4.2\times 10^{-13}~(90\%~\text{CL}).
\eeq

\subsection{FCNC Constraints in the Minimal Mixing Scenario}
Here, we investigate the FCNC constraints in the minimal mixing scenario. We search for the parameter space which is the same as the previous section in Fig.~\ref{fig:uniorg} and Fig.~\ref{fig:nonuniSorg}. In this scenario, the CKM matrix appearing in the Yukawa couplings in Eq.\,(\ref{eq:SCKM}) leads to the flavor mixing in the squark mass matrices at the weak scale due to the non-universality of the split family structure at the GUT scale.

Such flavor mixing is constrained by $\epsilon_K$, where the CP violation comes solely from the CP phase of the CKM matrix. In Fig.~\ref{fig:uni}, the red hatched regions are excluded. There, the region with the small $m_0$ and $M_{1/2}$ is excluded due to the light gluino and squarks. Notice that the SUSY FCNC contribution is larger for a larger $m_3^2/m_0^2$, and thus the constraint is severe for the left figure.%
\footnote{In our analysis, we fully diagonalize the squark masses without using the mass insertion technique (See appendix~\ref{sec:spliteffect} for more details).} In Fig.~\ref{fig:nonuniS}, on the other hand, no constraint appears from $\epsilon_K$. This is because the SUSY contributions are suppressed due to the heavy gluino and squarks for $|M_3|\simeq2.5-2.6$\,TeV. 
 
As a result, the Split-Family SUSY model is not stringently constrained by the flavor violation in the minimal mixing scenario. We have also confirmed that ${\mit  \Delta}  M_{K_{\rm SUSY}}$ and ${\mit  \Delta}  M_{D_{\rm SUSY}}$ become much smaller than the observed one (See $e.g.$ Tab.~\ref{tab:uni} and Tab.~\ref{tab:nonuni1}).%
\footnote{In the minimal mixing scenario, we have also confirmed that the neutron electric dipole moment (EDM) from the CKM phase does not lead to the stringent constraint. See also Ref.~\cite{Endo:2010fk} for the EDM constraints. }

Several comments are in order. As we have mentioned earlier, we may consider another simple family basis in Eq.\,(\ref{eq:SCKMu}). Since the flavor mixing is dominated by the renormalization group effects, we obtained similar constraint even in this case for the large ${\rm tan}\beta$. 

One may also wonder how large SUSY FCNC contributions are expected if the first and the second generation soft SUSY breaking masses are not degenerate. We discuss this possibility in appendix~\ref{sec:nondeg}. There, the model is severely constrained by the FCNC even for the minimal mixing scenario.

\subsection{FCNC Constraints in Small Mixing Scenario}
Now, let us discuss the FCNC problems in the small mixing scenario. As we will see, the flavor mixing in the slepton sector leads to sizable FCNC contributions because their masses are of $\mathcal{O}(100)$\,GeV to explain the muon $g-2$.

\subsubsection{FCNC Constraints on Slepton Flavor Mixing}
As we explained in Sec.~\ref{sec:model}, we take the diagonal lepton Yukawa coupling while allowing small flavor mixing in the SUSY breaking soft masses as in Eq.\,(\ref{eq:massmix}). We use the following parametrization for the slepton mixing matrix $V_{\rm mix}$,
\beq
\label{eq:leptonmix}
{\small
V_{\rm mix}=
\left(
\begin{array}{ccc}
1 & 0 & 0\\
0 & {\rm cos}\,\theta_{23} & {\rm sin}\,\theta_{23}\\
0 & -{\rm sin}\,\theta_{23} &  {\rm cos}\,\theta_{23}
\end{array}
\right)
\left(
\begin{array}{ccc}
 {\rm cos}\,\theta_{13} & 0 & {\rm sin}\,\theta_{13} \\
0 & 1 & 0\\
-{\rm sin}\,\theta_{13} & 0 &  {\rm cos}\,\theta_{13}
\end{array}
\right)
\left(
\begin{array}{ccc}
 {\rm cos}\,\theta_{12} & {\rm sin}\,\theta_{12} & 0\\
-{\rm sin}\,\theta_{12} &  {\rm cos}\,\theta_{12} & 0\\
0 & 0 & 1
\end{array}
\right).}
\eeq 
Here, we again assume that no CP violation appears in the mixing matrix. 

As the split-family structure is motivated by the Yukawa hierarchy, we also assume that the flavor mixing in the SUSY breaking sector has a structure similar to the CKM matrix. That is, we assume that the mixing angles are controlled by a small parameter  $\epsilon$, 
\beq
\label{eq:angles}
({\rm sin}\,\theta_{12},{\rm sin}\,\theta_{13},{\rm sin}\,\theta_{23})=(\mathcal{O}(\epsilon),\mathcal{O}(\epsilon^3),\mathcal{O}(\epsilon^2)),
\eeq
where $\epsilon \simeq  0.2$ roughly mimics the mixing angles in the CKM matrix. In the following, we demonstrate how large flavor mixing is allowed in the slepton sector by using $\epsilon$.

In our numerical analysis, we take the following mixing angles,
\beq
\label{eq:random}
(s_{12},s_{13},s_{23})=(R_1\epsilon,R_2\epsilon^3,R_3\epsilon^2),
\eeq
where $R_i$ shows the random numbers between $(0.5,~1.5)$. The random numbers $R_i$ for $m_{\tilde{L}}^2$ and $m_{\tilde{E}}^2$ are taken independently.

\begin{figure}[tbp]
\begin{center}
  \begin{minipage}{.4\linewidth}
  \includegraphics[width=\linewidth]{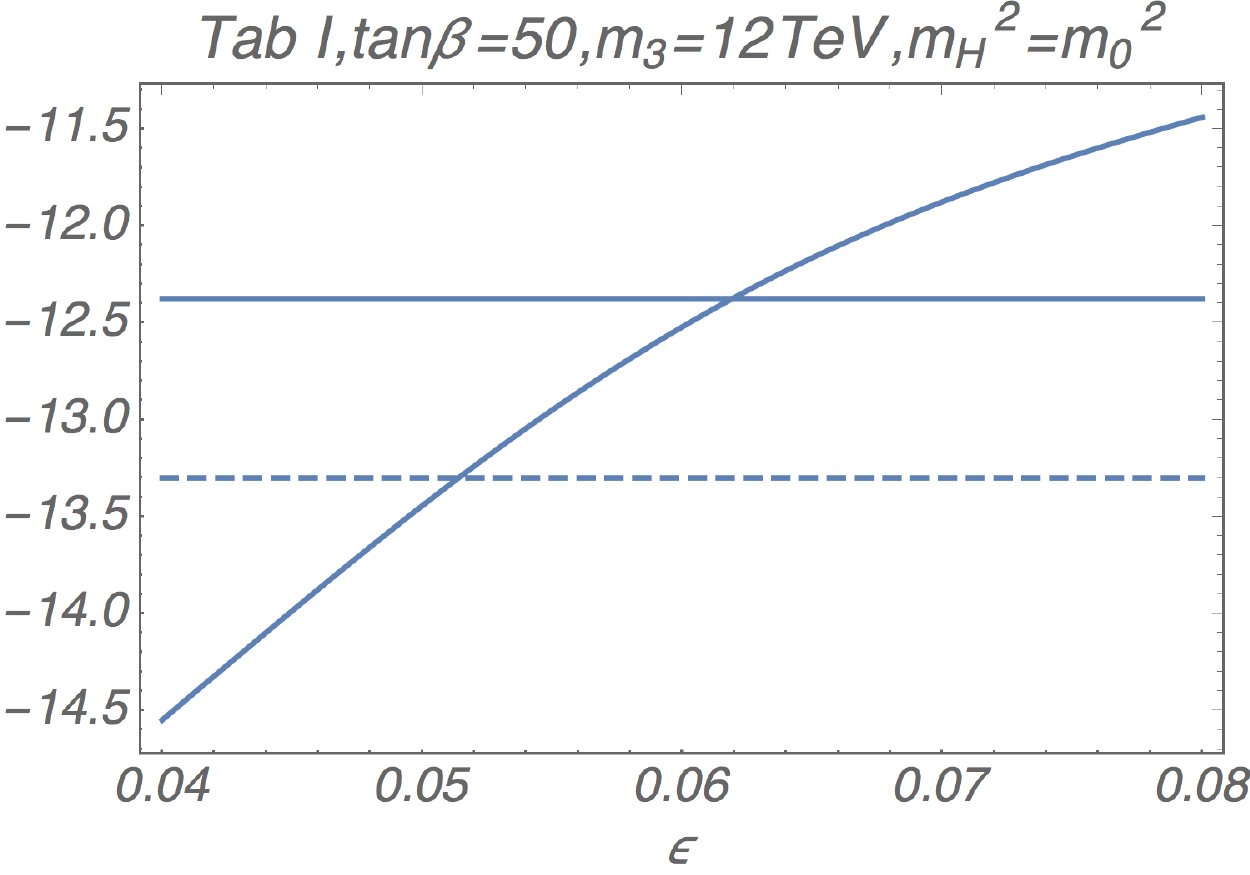}
 \end{minipage}
 \hspace{1cm} 
  \begin{minipage}{.4\linewidth}
  \includegraphics[width=\linewidth]{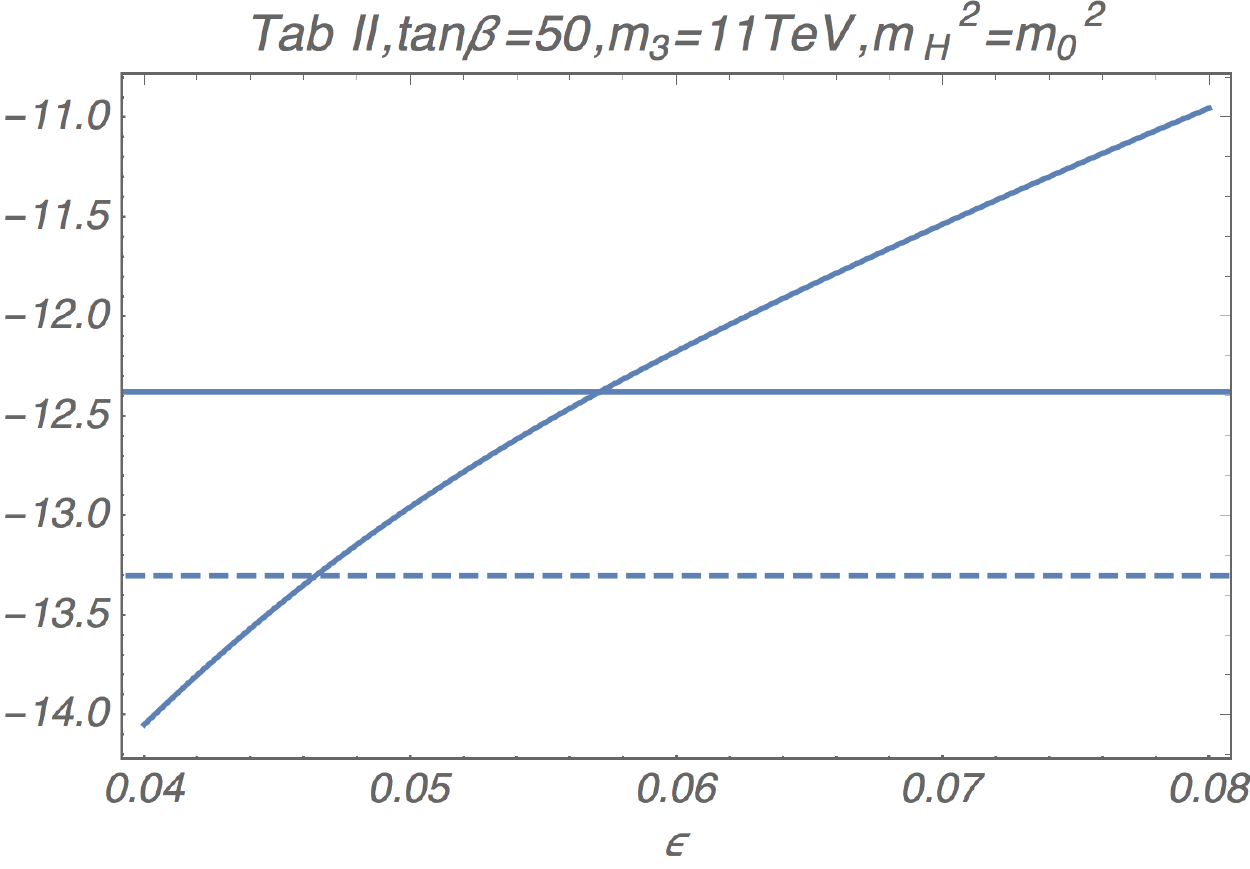}
 \end{minipage}
 \end{center}
  
\caption{\sl \small
The plot of $\mathcal{B}(\mu^+\rightarrow e^+\gamma)$ (the curve line) for given $\epsilon$. We used parameters in Tab.~\ref{tab:uni} and Tab.~\ref{tab:nonuni1} for the left side and the right side figures, respectively. The solid (dashed) horizontal lines are the current (future expected) upper-bound.
}
\label{fig:lfv}
\end{figure}

In Fig.~\ref{fig:lfv}, we show how large slepton mixing is tolerable from the current and future constraints on $\mathcal{B}(\mu^+\rightarrow e^+\gamma)$. There, we consider the input parameters in Tab.~\ref{tab:uni} and Tab.~\ref{tab:nonuni1} for examples.  The blue curve line denotes the model predictions for $\mathcal{B}(\mu^+\rightarrow e^+\gamma)$. The horizontal (dashed) lines are the current (future expected~\cite{Baldini:2018nnn,Hewett:2012ns}) experimental bounds. 
From the figures, the constraint from $\mathcal{B}(\mu^+\rightarrow e^+\gamma)$ requires $\epsilon\lesssim 0.06$.%
\footnote{We checked that the constraints are not significantly changed even if we put additional CP phases in $V_{\rm mix}$ of the sleptons. Detail analysis will be done elsewhere.}

Before closing this section, let us comment on the effect of the PMNS matrix. As in the case of the squark mixing, the PMNS matrix could lead to large flavor mixing in the slepton mass matrix at the weak scale even if we assume a diagonal soft masses  (See Eq.\,(\ref{eq:mass})). However, the FCNC constraints due to the PMNS matrix are not stringent if the neutrino Yukawa matrix is small enough as long as the right-handed neutrino mass $M_R\lesssim 10^{10}$\,GeV (See appendix~\ref{sec:lfvmns} for more details).

\subsubsection{FCNC Constraints on Squark Flavor Mixing}
In the small mixing scenario, we can also introduce $V_{\rm mix}$ to the squarks with the mixing angles given by Eq.\,(\ref{eq:angles}). In this case, however, the FCNC constraints on the mixing angle $\epsilon$ are much weaker than the slepton case, and hence, we do not discuss them any further.

\section{Bottom-tau unification}
\label{sec:btauuni} 
In the Split-Family SUSY model, the muon $g-2$ can be explained by the small $|M_1|,~|M_2|,~m_0^2$, and the large ${\rm tan}\beta$. In addition, the stringent limits from the LHC experiments require the rather large $|M_3|$. Interestingly, these parameter sets are found to be appropriate to realize the bottom-tau unification~\cite{Ajaib:2014ana}.

The bottom Yukawa coupling receives threshold corrections from the gluino loop diagrams~\cite{Banks:1987iu, Hall:1993gn, Hempfling:1993kv,Blazek:1995nv},
\beq
\label{eq:threshold}
\frac{{\mit  \Delta}  y_b}{y_b} \propto \frac{M_3\mu\, {\rm tan}\beta}{m_3^2},
\eeq
where $y_b$ is the bottom Yukawa coupling in the MSSM, and ${\mit  \Delta}  y_b$ is the threshold correction. The negative $M_3\, \mu\, {\rm tan}\beta$ gives the negative ${\mit  \Delta}  y_b$, which makes the bottom Yukawa coupling in the MSSM $y_b$ larger for a given bottom quark mass.%
\footnote{In our notation, $y_b$ is real positive (See Eq.\,(\ref{eq:SCKM})).}
For the tau Yukawa coupling, on the other hand, the threshold corrections are small due to the small $M_1$ and $M_2$ as well as the small gauge coupling constants. As a result, the large negative contribution in Eq.\,(\ref{eq:threshold}) makes $y_b/y_\tau$ large at the weak scale, which is appropriate for the bottom-tau unification at the GUT scale. This is the reason that we take negative $M_3$ in the previous section.

\begin{figure}[tbp]
\begin{center}
  \begin{minipage}{.4\linewidth}
  \includegraphics[width=\linewidth]{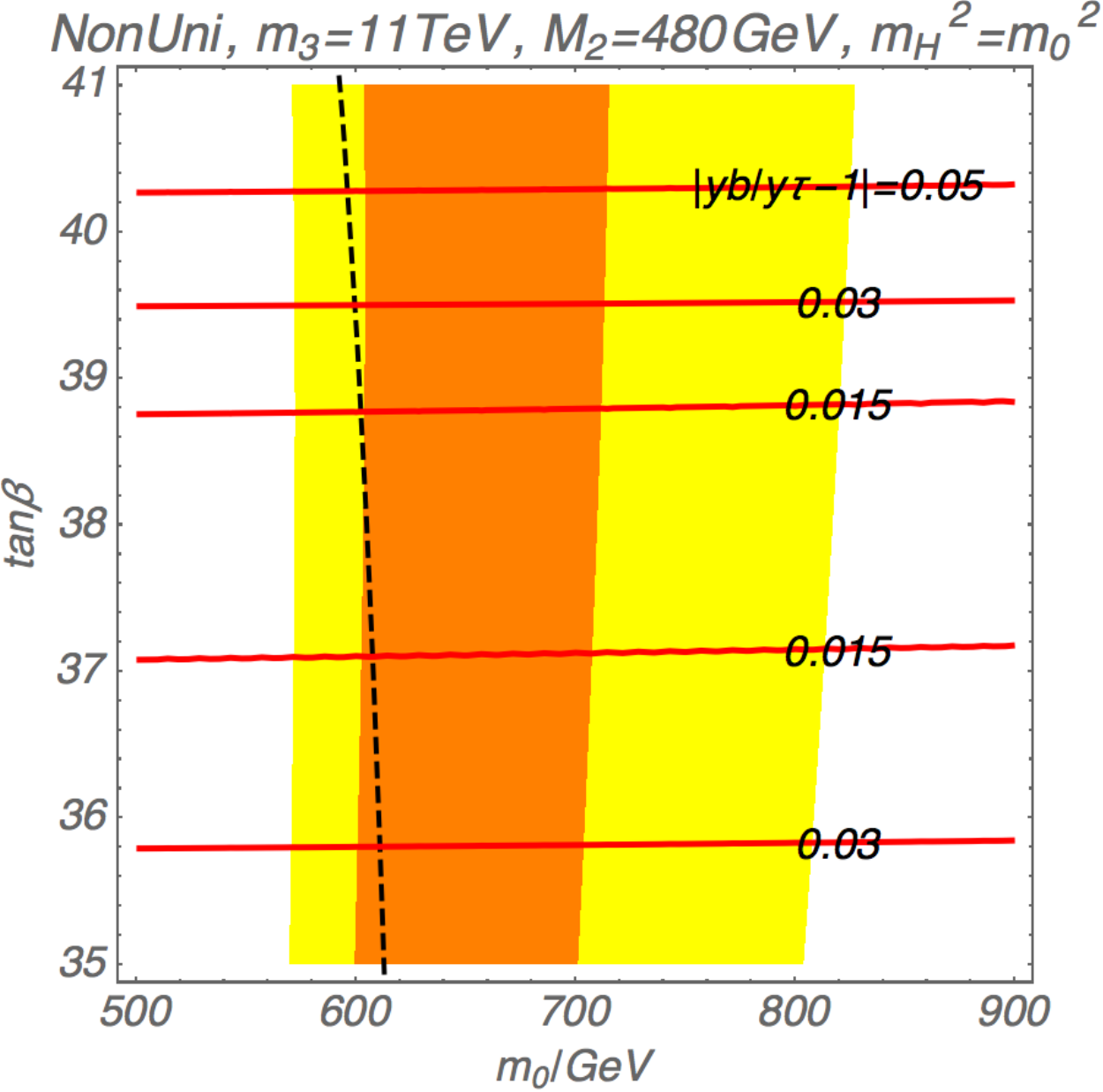}
 \end{minipage}
 \end{center}
\caption{\sl \small
The plot for the successful bottom-tau unification on the $(m_0-{\rm tan}\beta)$ plane. The red line shows the $|y_b/y_\tau-1|=0.015,~0.03,~0.05$. The color code is the same as Fig.~\ref{fig:nonuniS}. In most of the favored region in the figure, we have confirmed that the Higgs mass is in $124{\rm-}126$\,GeV and the dark matter abundance is $\Omega h^2 \simeq 0.12$ (the right side of the dotted line).}
\label{fig:nonunibt}
\end{figure}

In Fig.~\ref{fig:nonunibt}, we show the parameter space for the successful bottom-tau unification. There, we assume the input parameters in Tab.\,\ref{tab:nonuni1} other than $m_0$ and ${\rm tan}\beta$. The red lines are the contours of $|y_b/y_\tau-1|=~1.5\%,~3\%,~5\%$, where $y_b$ and $y_\tau$ are the bottom and tau Yukawa couplings at the GUT scale, respectively.%
\footnote{The threshold correction from the gluino is sensitive to the matching scale, which requires careful treatments of the decoupling and the renormalization group effects~\cite{Chigusa:2016ody}.}
 The other color codes are the same as Fig.~\ref{fig:nonuniS}. We have checked that the Higgs mass is in $124{\rm-}126$\,GeV in most of the favored region. We have also confirmed that the dark matter abundance is $\Omega h^2 \simeq 0.12$ on the right side of the dotted line (bino LSP region) in the figure.%
 \footnote{On the left side of the dotted line, the sneutrino is the LSP as discussed in Sec.~\ref{sec:pheno}. } Thus, we find the parameter space favored by the muon $g-2$ is also appropriate to achieve the bottom-tau unification.

\newpage

\section{Conclusion}
\label{sec:conclusion}
In this paper, we have revisited the Split-Family SUSY model. In the model, the sfermion masses of the first two generations are in the hundreds GeV range, while that of the third generation is in the tens TeV range. With this spectrum, the deviation of the muon $g-2$ and the observed Higgs boson mass are explained simultaneously.



In Sec.~\ref{sec:pheno}, we have first shown the parameter space to explain the muon $g-2$ and the Higgs boson mass. In our analysis, we have searched for two cases of the universal gaugino mass and the non-universal gaugino masses. For the universal gaugino mass, almost the entire region to explain the muon $g-2$ within $2\sigma$ is excluded by the collider searches as shown in Fig.~\ref{fig:uni}. This is due to the lightness of the squarks, the gluino, and the wino masses. For the non-universal gaugino masses, the gluino can be heavier with which the collider constraints can be easily evaded (See Fig.~\ref{fig:nonuniS}).%
\footnote{See $e.g.$ Ref.~\cite{Endo:2013lva,Endo:2013xka} for the discussion about the future sensitivity at the LHC and the ILC.}  We have also found the parameter space where the bino LSP can explain the observed dark matter density thanks to the coannihilation with the wino.


In Sec.~\ref{sec:fcnc}, we have studied the FCNC problems in the Split-Family SUSY model. We have searched for two scenarios, $i.e.$ the minimal mixing scenario and the small mixing scenario (See Sec.~\ref{sec:fcnc} for details). In the minimal scenario, we have assumed that the CKM matrix is the only source of the flavor mixing. There, we have shown that the SUSY FCNC contributions are small enough to evade the problem. 

For the small mixing scenario, we have assumed the CKM like mixing matrix to the soft mass parameters (See Eq.\,(\ref{eq:leptonmix}) and around it). Then, we have demonstrated how large flavor mixing is allowed in the slepton sector. There, the most stringent constraint comes from the lepton flavor violation decay of $\mu^+ \to e^++\gamma$. We have shown that the mixing angles have to be relatively small, $\epsilon\lesssim 0.06$.

In Sec.~\ref{sec:btauuni}, we have discussed one bonus feature of the model, the bottom-tau unification. For the successful bottom-tau unification, the large  threshold correction is required. Interestingly, such parameter space is compatible with the one favored by the muon $g-2$. In Fig.~\ref{fig:nonunibt}, we have shown that the bottom-tau unification is significantly improved for the large ${\rm tan}\beta$.

Several comments are in order. First, it is possible to achieve a small $\mu$ term in the Split-Family SUSY model. In fact, for $m_{H_{u,d}}^2=m_3^2$, the focus point mechanism~\cite{Feng:1999mn,Feng:1999zg} results in a small $\mu$ term. In such a case, the neutralino LSP can have a sizable Higgsino contribution, so that the dark matter-nucleon cross section becomes large. 

Throughout this paper, we have assumed that the SUSY breaking parameters do not have any CP-violating phases. In fact, these are strong assumptions and it is highly non-trivial to achieve such soft SUSY breaking parameters from high energy theory. The CP violating phases in the SUSY breaking parameters are constrained by the measurements of the EDMs, which will be discussed elsewhere.

\begin{acknowledgments}
This work is supported by JSPS KAKENHI Grant Numbers JP26104001 (T.T.Y), JP26104009 (T.T.Y), JP16H02176 (T.T.Y), JP17H02878 (M. I. and T. T. Y.), No. 15H05889 and No. 16H03991 (M. I.), JP15H05889 (N.Y.), JP15K21733 (N.Y.), JP17H05396 (N.Y.), JP17H02875 (N.Y.), and by World Premier International Research Center Initiative (WPI Initiative), MEXT, Japan (T.T.Y.). The work of M. S. is supported in part by a Research Fellowship for Young Scientists from the Japan Society for the Promotion of Science (JSPS).
\end{acknowledgments}

\appendix
\section{Validity of the Mass Insertion Approximation}
\label{sec:spliteffect}
In this appendix, we briefly discuss the validity of the mass insertion approximation for the SUSY FCNC contributions. For simplicity, we consider the case of the universal gaugino mass in the minimal mixing scenario. 

In the mass insertion approximation, the gluino contribution to $\epsilon_K$ is estimated at the next leading order~\cite{Ciuchini:1998ix},
\beq
\label{eq:epsilon}
\epsilon_{K_{\rm SUSY}} &\simeq& -\frac{3}{20\sqrt{2}{\mit  \Delta}  M_K^{\rm exp}}M_K f_K^2\frac{\alpha_s^2}{216 m_0^2}(24 x f_6(x)+66 \tilde{f}_6(x)){\rm Im}[({\mit  \Delta} ^{ds}_{LL})^2].
\eeq
Here, $\alpha_s=0.1184$, $m_K=0.498$\,GeV, the Kaon decay constant $f_K=0.16$\,GeV. The mixing parameter ${\mit  \Delta} ^{ds}_{LL}$ denotes an off-diagonal element for the left-handed down and strange squarks soft SUSY breaking  mass squared matrix normalized by $1/m_{0}^2$,
\beq
\label{eq:mix}
{\mit  \Delta} ^{ds}_{LL}&\simeq& V_{ts}^{\rm CKM}V_{td}^{*{\rm CKM}}\left(\frac{m_3^2}{m_{0}^2}\right).
\eeq
From Eq.\,(\ref{eq:epsilon}),\,(\ref{eq:mix}), $\epsilon_{K_{SUSY}}$ is enhanced by a factor of $m_3^4$. 

The above approximation is, however, only valid when the sfermion mass splitting is small, $i.e.$ $m_0^2\simeq m_3^2$. For $ m_3^2\gg m_0^2$, on the other hand, the approximation leads to the overestimation. Thus, we need to calculate the FCNC processes in the exact mass diagonalization. For this purpose, we use the package {\tt susy\_flavor\_2\_54}~\cite{Rosiek:2014sia,Crivellin:2012jv,Rosiek:2010ug}.

\begin{figure}[tbp]
\begin{center}
 \begin{minipage}{.6\linewidth}
  \includegraphics[width=\linewidth]{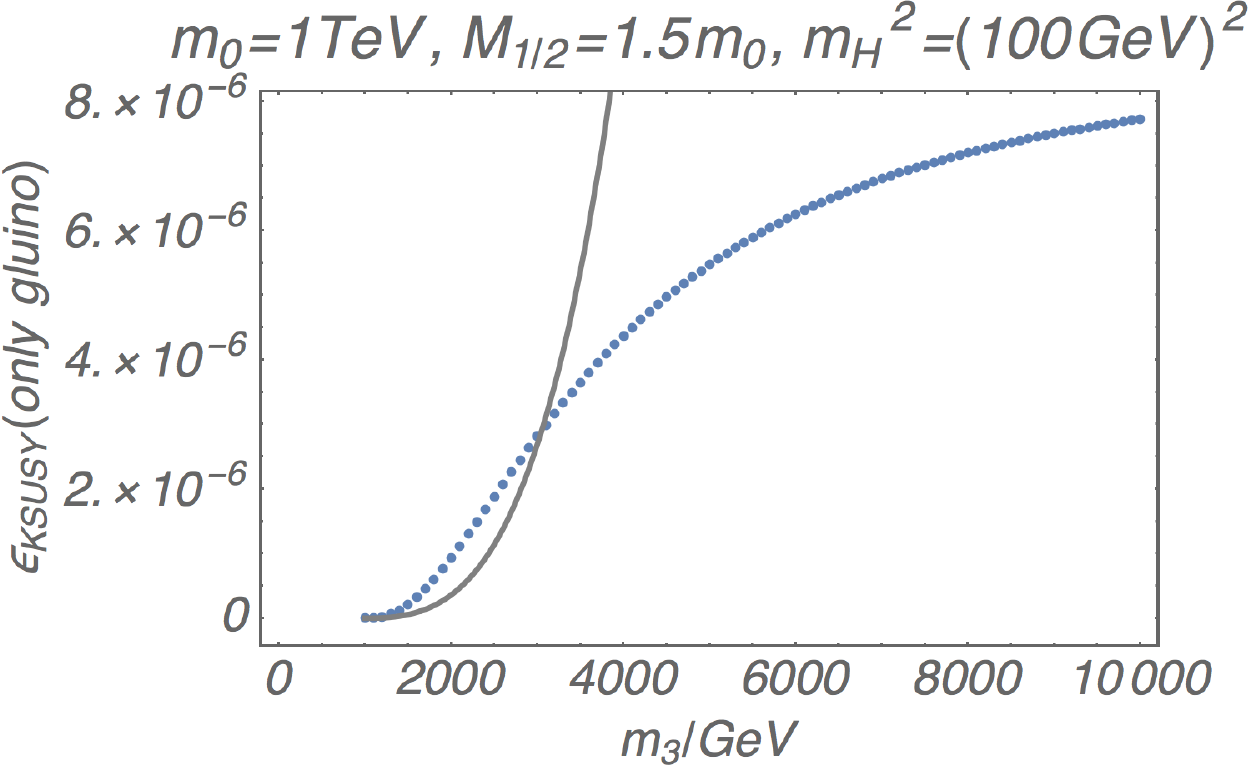}
 \end{minipage}
 \end{center}
  \caption{\sl \small The plot of the gluino contribution to  $\epsilon_K$ for a function of $m_3$. The blue dotted line shows the numerical result by the only {\tt susy\_flavor}. The gray line denotes the result by the mass insertion approximation in Eq.\,(\ref{eq:epsilon}). Note that all calculations include only so-called $VLL$ element~\cite{Crivellin:2012jv} to compare with the mass insertion approximation. 
}
\label{fig:plotenhance}
\end{figure}

 The numerical result of $\epsilon_{K_{\rm SUSY}}$ as a function of $m_3$ is shown in Fig.~\ref{fig:plotenhance}.  The blue dotted line corresponds to the result from the code, where we take the input parameter $m_0=1\,{\rm TeV},~M_{1/2}=1.5\,{\rm TeV},~{\rm tan}\beta=40,~m_{H_{u,d}}^2=(100\,{\rm GeV})^2$.%
 \footnote{Here, we only include the contribution from the $VLL$ four-quark operator~\cite{Crivellin:2012jv} to compare with the approximation in Eq.\,(\ref{eq:epsilon}). We also switched off the resummation of chirally enhanced corrections~\cite{Crivellin:2012jv} to compare with the mass insertion approximation. }  The gray solid line is given by the mass insertion approximation in Eq.\,(\ref{eq:epsilon}) for the same input parameters. 
Here, these input parameters are set at the SUSY scale to make comparison easier. From the figure, we confirm the overestimation of $\epsilon_K$ for $m_3^2\gg m_0^2$ in the mass insertion approximation.

\section{FCNC Constraints on First and Second Sfermion Soft Mass Splitting}
\label{sec:nondeg}

In this appendix, we discuss how large mass splitting is allowed for the first and the second generation squarks from the FCNC constraints. To parametrize the splitting, we use the following non-universal soft masses for the squarks,
\beq
m_{\rm soft}^2=
\left(
\begin{array}{ccc}
m_0^2 & 0 & 0 \\
0 & m_2^2 & 0 \\
0 & 0 & m_3^2 
\end{array}
\right),
\eeq
where $m_2$ denotes the soft mass parameter of the second generation squarks. 

\begin{figure}[tbp]
\begin{center}
 \begin{minipage}{.4\linewidth}
  \includegraphics[width=\linewidth]{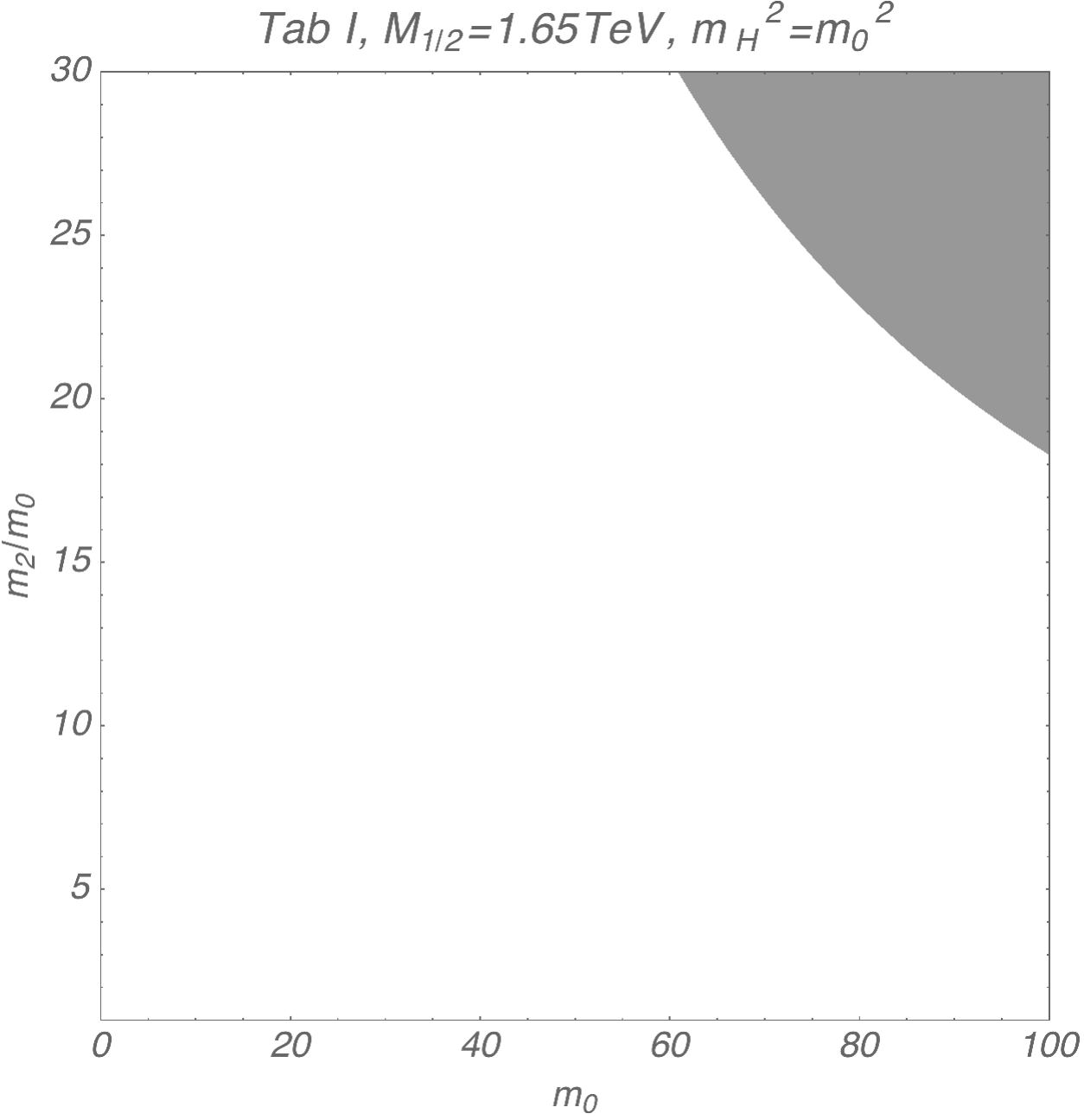}
 \end{minipage}
\hspace{1cm} 
   \begin{minipage}{.4\linewidth}
  \includegraphics[width=\linewidth]{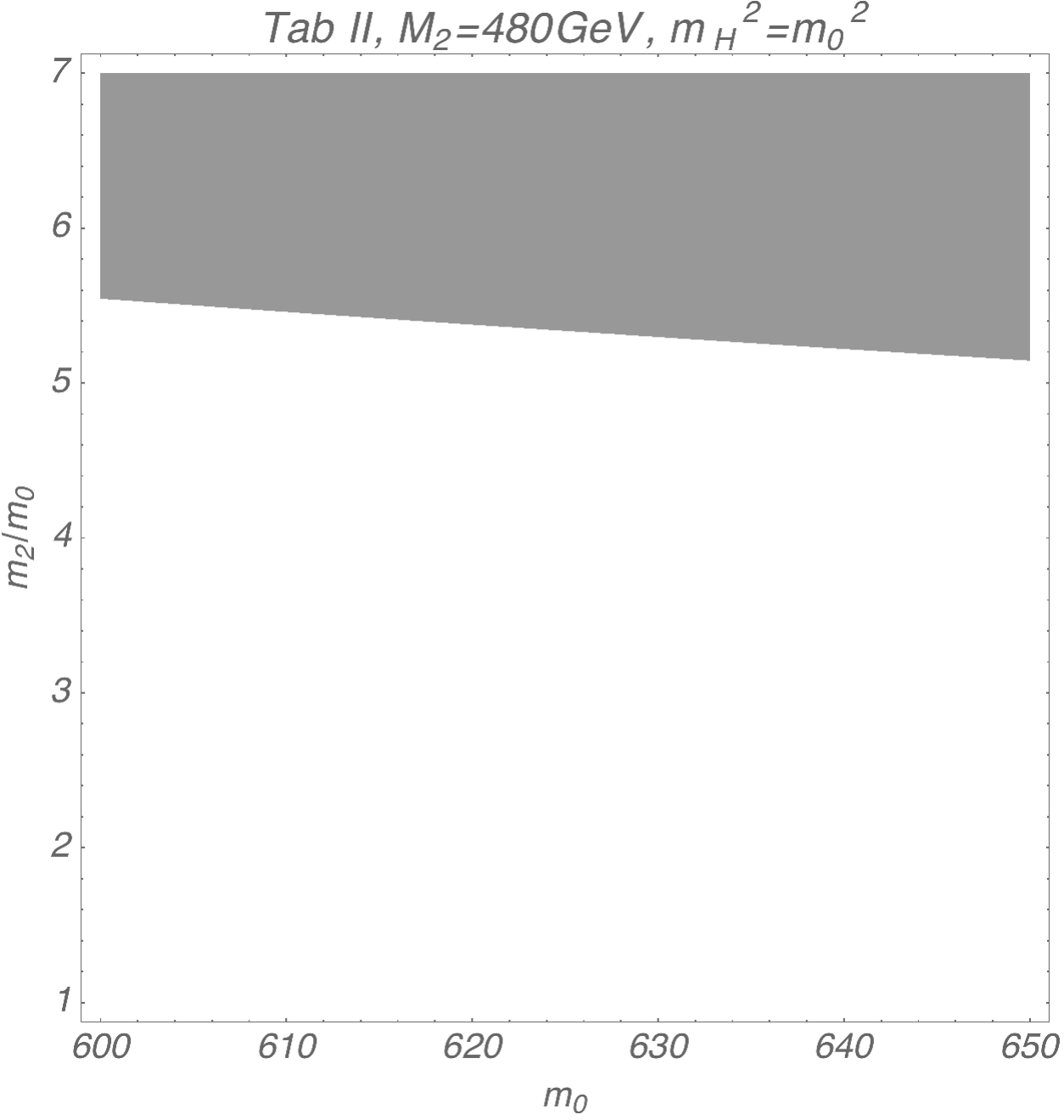}
 \end{minipage}
 \end{center}
  
\caption{\sl \small The FCNC constraints for the mass splitting between the first two generation squarks. The gray shaded regions are excluded by $\epsilon_K$ ($90$\%\,CL).
}
\label{fig:constratio}
\end{figure}

In Fig.~\ref{fig:constratio}, we show the FCNC constraints for the case of the universal gaugino mass (the figure in the left). The gray shaded region is excluded by $\epsilon_K$ ($90\%$\,CL). There, we use the input parameters in Tab.~\ref{tab:uni}, but take $m_0$ and $m_2$ as variables. The parameter space with the ratio $m_2/m_0= \mathcal{O}(10)$ is excluded for $m_0\gtrsim 60$\,GeV, where $m_2$ becomes comparable to the gluino mass contributions to the squark masses in the second generation.

In the right panel of Fig.~\ref{fig:constratio}, we also show the constraints for the non-universal gaugino masses. There, we use the input parameters in Tab.~\ref{tab:nonuni1}. The SUSY FCNC contributions are  large with the splitting, $m_2/m_1\gtrsim 5$, where $m_2$ becomes comparable to the gluino mass contributions to the squark masses in the second generation.

\section{Lepton Flavor Violation from MNS Matrix Effect}
\label{sec:lfvmns}
In this appendix, we briefly discuss the constraint from the decay $\mu^+\to e^++\gamma$, where the MNS matrix is taken into account as in the case of the minimal mixing scenario in the squark sector.

In the presence of the right-handed neutrinos, the superpotential of the lepton sector is given by,
\beq
W=\hat{f}_E^{ij}L^{[e]}_i\bar{E}^{[e]}_j H_d+(U_{\rm MNS}^*\hat{f}_{\nu})^{ij}L^{[e]}_i N_{Rj} H_u+\frac{1}{2}M_RN_{Ri}N_{Ri}.
\eeq
Here, $N_{Ri}\,(i=1{\rm-}3)$ are the right-handed neutrinos with three flavors. As in the case of the minimal mixing scenario in the squark sector, we assume the soft masses in Eq.\,(\ref{eq:mass}) while the Yukawa couplings are diagonal matrices ($\hat{f}_\nu$) with the MNS matrix ($U_{\rm MNS}$).

In this setup, the flavor mixing parameter between the first and second generation of the charged selectrons  is given by,
\beq
({\mit  \Delta} ^e_{12})_{LL}\simeq \frac{1}{16\pi^2}\left(\sqrt{\frac{M_R}{10^{15}\,{\rm GeV}}}\right)^2\left(\frac{m_3^2}{m_0^2}\right)(U_{\rm MNS})_{13}(U_{\rm MNS}^*)_{23} {\rm ln}(M_R/10^{16}\,{\rm GeV}),
\eeq
where the parameter is defined by an off-diagonal element for the left-handed electron and muon sleptons soft SUSY breaking mass squared matrix normalized by $1/m_0^2$.

Then, the branching ratio of the process $\mu^+\to e^++\gamma$ is given in the following~\cite{Hisano:1995cp,Arganda:2005ji,Paradisi:2005fk},
\beq
\mathcal{B}\simeq \frac{48 \pi^3 \alpha_{\rm em}}{G_F^2}\left(c_1\frac{\alpha_1}{4\pi}\frac{\mu\,m_{\rm bino}{\rm tan}\beta}{m_{\rm slepton}^4}({\mit  \Delta} ^e_{12})_{LL}\right)^2.
\eeq
Here, $G_F=1.166 \times 10^{-5}\,{\rm GeV}^{-2}$ is the Fermi coupling decay constant, $\alpha_{\rm em}\simeq 1/137$ is the fine-structure constant, $\alpha_1\simeq 1/60$ is the weak coupling constant. We also take $m_{\rm slepton}\simeq 480$\,GeV, $\mu\simeq9\times 10^3$\,GeV, and $m_{\rm bino}\simeq400$\,GeV as in Tab.~\ref{tab:nonuni1}. The coefficient $c_1$ is about $0.1$~\cite{Hisano:1995cp,Arganda:2005ji,Paradisi:2005fk}. The branching ratio is consistent with the current upper-limit $\mathcal{B}\lesssim 10^{-13}$ for $M_R\lesssim 10^{10}\,{\rm GeV}$. More detailed analysis remains for future work.

\bibliography{../papers2}

\end{document}